\documentclass[12pt]{article}
\usepackage{mathrsfs}
\usepackage{graphicx}
\usepackage{amssymb,amsmath,amsthm}
\usepackage{bm}
\usepackage[pdftex,urlcolor=blue]{hyperref}
 \textwidth = 6.5 in \textheight = 9 in \oddsidemargin =
0.0 in \evensidemargin = 0.0 in \topmargin = 0.0 in \headheight =
0.0 in \headsep = 0.0 in
\def\ba{\begin{eqnarray}}
\def\ea{\end{eqnarray}}
\def\nn{\nonumber}

\def\H{\mathcal{H}}

\def\P{\mathcal{P}}

\def\O{\mathcal{O}}

\def\V{\mathcal{V}}
\def\:{\boldsymbol{:}}

\def\dk#1{\stackrel{\circ}{#1}}
\def\dkt#1{\stackrel{\bullet}{#1}}
\def\k{\mathbf{k}}
\def\F{\mathcal{F}}

\long\def\symbolfootnote[#1]#2{\begingroup%
\def\thefootnote{\fnsymbol{footnote}}\footnote[#1]{#2}\endgroup}

\begin{document}

\begin{center} 
\vglue .06in
{\Large \bf {Reconstructing Single Field Inflationary Actions From CMBR Data.\symbolfootnote[1]{This work was supported by the US department of energy.}}}\\[.5in]

{\bf Christopher S. Gauthier\symbolfootnote[2]{Electronic address: csg@umich.edu} and Ratindranath Akhoury\symbolfootnote[3]{Electronic address: akhoury@umich.edu}}\\
[.1in]
{\it{Michigan Center for Theoretical Physics\\
Randall Laboratory of Physics\\
University of Michigan\\
Ann Arbor, Michigan 48109-1120, USA}}\\[.2in]
\end{center}

\begin{abstract}
This paper describes a general program for deriving the action of single field inflation models with nonstandard kinetic energy terms using CMBR power spectrum data. This method assumes that an action depends on a set of undetermined functions, each of which is a function of either the inflaton wave function or its time derivative. The scalar, tensor and non-gaussianity of the curvature perturbation spectrum are used to derive a set of reconstruction equations whose solution set can specify up to three of the undetermined functions. The method is then used to find the undetermined functions in various types of actions assuming power law type scalar and tensor spectra. In actions that contain only two unknown functions, the third reconstruction equation implies a consistency relation between the non-gaussianty, sound speed and slow roll parameters. In particular we focus on reconstructing a generalized DBI action with an unknown potential and warp factor. We find that for realistic scalar and tensor spectra, the reconstructed warp factor and potential are very similar to the theoretically derived result. Furthermore, physical consistency of the reconstructed warp factor and potential imposes strict constraints on the scalar and tensor spectral indices.
\end{abstract}

\section{Introduction}
Since the landmark COBE experiment, the study of the cosmos has entered a new age of precision cosmology. For the first time in the history of modern cosmology, direct quantitative measurements of early cosmological observables were available. The data taken by COBE was critical in establishing inflation as the central paradigm in our theories of the origin of the universe \cite{Smoot:1992td}. Thanks to experiments that measured the spectrum of CMBR fluctuations, we have now confirmed that the near-scale invariance of large scale fluctuations that is a prediction of inflation are in fact borne out in the data. Although observation supports the general theory of inflation, as of now the data is unable to conclusively determine the mechanism responsible for inflation. 

The difficultly in discriminating between different inflation models lies in fact that all current models of inflation predict the same near-scale invariant spectrum of scalar fluctuations. To further narrow down the number of observationally consistent inflation models, observables independent of the scalar perturbation need to be measured and compared to model predictions. Two additional inflationary observables are the spectrum of tensor perturbations $P_{h}$ \cite{Grishchuk:1974ny,Rubakov:1982df} and the non-gaussianity $f_{NL}$ \cite{Gangui:1993tt,Maldacena:2002vr,Chen:2006nt} of the CMBR temperature perturbation spectrum. In recent years greater progress has been made in measuring these quantities directly. Upper bounds on the amplitude of the tensor perturbation spectrum, which is the spectrum of relic gravitational waves, have been determined directly through analysis of the CMBR polarization \cite{Page:2006hz,Komatsu:2008hk}. The non-gaussianity, which represents the deviation of the curvature perturbation from gaussian statistics, is also being better understood. Analysis of WMAP3 data \cite{Yadav:2007yy} has found evidence of non-gaussian statistics in the CMBR temperature spectrum. With a better knowledge of these extra observables it becomes possible to better determine which model of inflation is most likely to have taken place. For example, a large $f_{NL}$ would tend to rule out a single field inflation model with minimal kinetic terms, while favoring those models that predict a large non-gaussianity.  

Ultimately, one would like to use the features of the CMB temperature anisotropy to reconstruct the inflaton action directly. It is customary to write the general scalar field action as  \cite{Garriga:1999vw}
\begin{gather}
S = \int d^{4}x \sqrt{-g} \, p(\phi,X)
,
\qquad
X = \frac{1}{2} g_{\mu\nu} \partial^{\mu} \phi \partial^{\nu} \phi.
\label{GeneralAction}
\end{gather}
Note that the Lagrangian (density) $p$ is exactly equal to the pressure, which is the motivation behind denoting the Lagrangian by $p$. In our analysis we will limit ourselves to actions that contain no third or higher derivatives of the inflaton. Throughout this paper we will assume that the curvature of the 3 non-compact space dimensions will be zero. Following the cosmological principle, our only choice for the metric is the FRW metric: $g^{\mu\nu} = \textrm{diag}(1,-a^{2},-a^{2},-a^{2})$, where $a$ is a time dependent scale factor. Using (\ref{GeneralAction}), the Friedmann equations for the scale factor are
\begin{gather}
3 M_{pl}^{2} H^{2}
=
\rho,
\label{FriedmannOne}
\\
-
2 M_{pl}^{2}
\dot{H}
=
\rho + p,
\label{FriedmannTwo}
\end{gather}
where $H = \frac{d \log a}{dt}$ is the Hubble parameter and $\rho$ is the energy density, which in terms of the Lagrangian is
\begin{gather}
\rho = 2 X p_{,X} - p.
\label{GenEnergyDensityOne}
\end{gather}
In single field inflation models with a minimal kinetic term the action is
\begin{gather}
S = \int d^{4}x \sqrt{-g} \left[X - V(\phi)\right].
\label{MinimalAction}
\end{gather}
If we assume (\ref{MinimalAction}), the only function that needs to be determined from the data is the potential $V(\phi)$. Reconstruction of the inflationary potential for models of the form (\ref{MinimalAction}) has been studied extensively \cite{Copeland:1993zn,Kolb:1994eu,Copeland:1993jj,Lidsey:1995np,Lesgourgues:2007gp,Lesgourgues:2007aa,Hamann:2008pb}. However, by assuming that the action has a minimal kinetic term we neglect a rich class of models such as DBI inflation \cite{Silverstein:2003hf,Chen:2004gc}, k-inflation \cite{ArmendarizPicon:1999rj} and ghost inflation \cite{ArkaniHamed:2003uz}. In contrast, only a hand full of articles have been written that deal with the reconstruction of inflationary action with general kinetic terms \cite{Peiris:2007gz,Bean:2008ga,Li:2008qc}. 

In non-minimal kinetic models the speed at which scalar fluctuations propagate can be different than the speed of light. This can effect the temperature anisotropy in two ways. First, if $c_{s} < c =1$, scalar fluctuations have a sound horizon that is smaller than the cosmological horizon, causing curvature perturbations to freeze in earlier than normal. Depending on how the Hubble parameter and sound speed change during the course of inflation, the temperature anisotropy can develop noticeable signatures of non-minimal kinetic terms.  Second, models with non-minimal kinetic terms will in general produce a non-gaussian spectrum. Traditionally, the non-gaussianity is measured by the non-linearity parameter $f_{NL}$ defined by the following ansatz for the curvature perturbation:
\begin{gather}
\zeta = \zeta_{L} - \frac{3}{5} f_{NL} \zeta_{L}^{2}.
\label{FnlDefinition}
\end{gather}
Here, $\zeta$ is the general curvature perturbation and $\zeta_{L}$ is a curvature perturbation with gaussian statistics. Within the standard canonical action (\ref{MinimalAction}), non-gaussianities can be produced by cubic or higher order terms in the inflaton potential or by secondary interactions with gravity \cite{Maldacena:2002vr}. However, non-gaussianities produced by these mechanisms are on the order of the slow roll parameters, and thus small. In contrast, models with non-minimal kinetic terms can have large non-gaussianities, providing a clear distinction from canonical inflation. 

The goal of this paper will be to reconstruct an inflationary action from observables starting with as few initial assumptions as possible. In this paper we take the experimental inputs to be the scalar curvature perturbation $P_{s}$, the tensor curvature perturbation $P_{h}$, and the non-gaussianity (non-linearity) parameter $f_{NL}$. Unfortunately, completely reconstructing the off-shell action is not possible since the observables only carry on-shell information. To understand why the off-shell action is inaccessible to us, consider the interpretation of the action $p(\phi,X)$ as a surface in the three dimensional space $(\phi,X,p)$ \cite{Bean:2008ga}. Because the observables are insensistive to the off-shell behavior of the action, we can only determine the one-dimensional trajectory  $p = p(\phi,X(\phi))$ of the action on-shell, embedded in the two-dimensional surface defined by $p = p(\phi,X)$. A one-dimensional trajectory has an infinite number of surfaces that contain it, each related to one another by a canonical transformation \cite{Bean:2008ga}. Therefore we have to be more specific about the form of the action that we are trying to find. In this paper we will reconstruct inflationary actions that have the form 
\begin{gather}
p(\phi , X)
=
P(g_{1}(X),...,g_{m}(X),f_{1}(\phi),...,f_{n}(\phi)).
\label{SeocndGeneralFormofAction}
\end{gather}
Here it is assumed that $P(x_{1},...,x_{m},y_{1},...,y_{n})$ is a known function of the $\{x_{i}\}$
and $\{y_{\alpha}\}$, and the functions $\{g_{i}\}$ and $\{f_{\alpha}\}$ are not all known. Once the on-shell trajectory $\phi=\phi(k)$ is determined, the action (\ref{SeocndGeneralFormofAction}) defines a surface in the $(\phi,X,p)$-space up to a field redefinition. 

The idea will be to use data on the CMBR perturbation spectrum to find the functions $\{g_{i}(X)\}$ and $\{f_{\alpha}(\phi)\}$. In a naive comparison with algebraic linear equations, we expect that if there are $n$ unknown functions, finding the action requires $n$ experimental inputs. Since we are assuming that there only three observables: $P_{s}$, $P_{h}$ and $f_{NL}$, we can derive three reconstruction equations, which can determine an action with three or fewer unknown functions. In the case where the number of unknown functions is less than the number of experimental inputs, the reconstruction equations not used to find the action become constraint equations. 

Since we are interested in solving for functions  $\{g_{i}\}$ and $\{f_{\alpha}\}$ and not just numbers, we will need to know at least a portion of $P_{s}$, $P_{h}$ and $f_{NL}$\footnote{In this paper $f_{NL}$ represents the equilateral bispectrum, and is therefore a function of a single scale.} as functions of the scale $k$. While the scale dependence of $P_{s}$ is known to be at least approximately power law dependent on $k$, the scale dependence of the other two observables $P_{h}$ and $f_{NL}$ is at this point unclear. Although future experiments will be able to clarify some aspects of the tensor and non-gaussianity signals, their exact functional forms will probably not be available for quite some time if at all. Regardless, the method we develop here does have utility outside of reconstruction. This method is well suited to testing how the form of an action depends on the observables. For instance if the scalar perturbation is of the near scale invariant variety:
\begin{gather}
P_{s}
\propto
k^{n_{s} - 1}
\label{PscalarFirst}
\end{gather}
we can use $P_{s}$ to help derive an action and study its dependence on the index $n_{s}$. That way if we wish to connect the action derived from (\ref{PscalarFirst}) to an action derived from theory, we can see if the theoretical action leads to reasonable results for the observables. Furthermore, as we mentioned earlier when there are only one or two unknown functions in (\ref{SeocndGeneralFormofAction}), the remaining reconstruction equations determine new consistency relations. In this paper most of the examples we deal with have only two unknown functions, which we solve for using the scalar and tensor spectrum data. The reconstruction equation derived from the non-gaussianity will then be a constraint; relating the non-gaussianity to the sound speed, the Hubble parameter and/or their derivatives. Outside of deriving the action we also find a method for quickly obtaining the sound speed as a function of time from the scalar and tensor perturbation spectra. Finding a sound speed different from the speed of light even over a small range of scales would be a powerful indication of non-canonical inflation.

This paper is organized as follows. In section \ref{EqusofInflation} we present the method for reconstructing the action from the scalar, tensor perturbation and the non-gaussianity parameter. We explain how cosmological data can find the Hubble parameter $H$, and the sound speed $c_{s}$, and how these in turn can be used to find three unknown functions of the action (\ref{SeocndGeneralFormofAction}). Once the method has been explained in section \ref{Examples} we carry out a derivation of the action for different functions $P(z_{1},z_{2},z_{3})$, assuming that the scalar and tensor power spectra both scale like $k$ to some power. In section \ref{DBIInflation} we apply our method to find the warp factor and potential in a generalized DBI inflation model. We find the warp factor and potential as functions of the spectral indices and the initial value of the Hubble parameter. The results for these are compared to the theoretically motivated warp factor and potential used in D3-brane DBI inflation. Finally, in section \ref{Conclusion} we review our main results.

\section{The Reconstruction Equations}
\label{EqusofInflation}
We start our derivation of the reconstruction procedure by explaining how the observables are used to find the Hubble parameter $H$ and sound speed $c_{s}$\footnote{The method described here was inspired by the technique used in \cite{Cline:2006db}}. Once we have these, the action can be obtained using a set of reconstruction equations that will be shown later. Let us begin by recalling the definition of the slow roll parameter\footnote{The term ``slow roll parameter'' is taken from chaotic inflation where inflation occurs only when the inflaton ``velocity'' $\dot{\phi}$ is small. However, DBI inflation can still occur for large $\dot{\phi}$.} $\epsilon$ in terms of the Hubble parameter. The definition of $\epsilon$ implies that
\begin{gather}
\frac{d H}{dt}
=
- \epsilon H^{2}.
\label{phiHEqu}
\end{gather}
Since the perturbation spectra and non-gaussianity are functions of $k$ and not time, we wish to rewrite this equation for $\frac{d H}{dt}$ into an equation for $\dk{H} = \frac{d H}{ d \log k}$. However, because we are assuming a general sound speed, we must be careful to differentiate between the horizons of scalar and tensor fluctuations. If the sound speed differs from unity (in particular $c_{s} < 1$), then the horizon size of scalar fluctuations: $(a H /c_{s})^{-1}$ is smaller than that of the tensor fluctuations: $(aH)^{-1}$. This implies that at any given time, the scales $k_{s}$ and $k_{t}$ at which the scalar and tensor fluctuations leave their respective horizons, will in general be different. For our purposes, we choose to study the dependence of $H$ on the scalar wave number $k_{s}$. Therefore, the condition for horizon exit is now $k c_{s} = a H$ instead of the more familiar relation: $k = a H$. Having made clear our choice of wave number, we now set out to express $\frac{d \log k}{dt}$ in terms of familiar quantities:
\begin{gather}
\frac{d \log k}{dt}
=
H 
(1 
- 
\epsilon 
+
\frac{\kappa}{H} 
\frac{d \log k}{dt}
)
\end{gather}
where we have defined $\kappa = - \frac{\dk{c_{s}}}{c_{s}}$. Solving for $\frac{d \log k}{dt}$ above:
\begin{gather}
\frac{d \log k}{dt}
=
\frac{H (1 - \epsilon)}{1 - \kappa}.
\label{dlogkdt}
\end{gather}
With equation (\ref{dlogkdt}) in hand, the equation for $\dk{H}$ can now be found from (\ref{phiHEqu}): 
\begin{gather}
\dk{H}
=
-
\frac{H \epsilon (1 - \kappa)}{1 - \epsilon}.
\label{DotHEqu}
\end{gather}
We will use this equation to find $c_{s}$ once we have found $H$ and $\epsilon$ in terms of the observables. Since $\epsilon$ depends on the time derivative of the Hubble parameter, $H(k)$ and $\epsilon(k)$ are independent parameters. Since we have two independent parameters, we will likely need two independent observables. The two observables we will use here are the scalar and tensor perturbation spectra. Recall that to first order in the slow roll parameters the perturbation spectra are given by 
\begin{gather}
P_{s}(k_{s})
=
\left.
\frac{ H^{2}}{8 \pi^{2} M_{pl}^{2} \epsilon c_{s}}
\right|_{k_{s} c_{s} = a H},
\label{Pscalar}
\end{gather}
\begin{gather}
P_{h}(k_{t})
=
\left.
\frac{2}{\pi^{2}}
\frac{H^{2}}{M_{pl}^{2}}
\right|_{k_{t} = a H}.
\label{PhspecEquDBI}
\end{gather}
The extra source of information can also be garnered from the non-gaussianity parameter $f_{NL}$. However, going in this route would result in a more complicated solution. The parameter $\epsilon$ can be found as a function of wave number using (\ref{Pscalar}). Solving for $\epsilon$ we obtain
\begin{gather}
\left.
\epsilon
\right|_{k_{s} c_{s} = a H}
=
\frac{1}{8 \pi^{2} M_{pl}^{2} P_{s}(k_{s})}
\left.
\frac{H^{2}}{c_{s}}
\right|_{k_{s} c_{s} = a H}.
\label{epsilonInTermsHandP}
\end{gather}
As a matter of convenience define $P_{s}  = A \P_{s}$, where $A$ is the value of the scalar perturbation at some fiducial scale $k = k_{0}$. If $k_{0} \simeq 0.002$Mpc${}^{-1}$ then present data suggests that $A \simeq 10^{-9}$. Here, $\P_{s}$ is the normalized scalar perturbation defined such that $\P_{s}(k_{0}) = 1$. Furthermore, let  $H = \alpha \H$ where $\alpha^{2} = 8 \pi^{2} M_{pl}^{2} A$. Substituting (\ref{epsilonInTermsHandP}) in for $\epsilon$ in equation (\ref{DotHEqu}) we have
\begin{gather}
\dk{\H}|_{k_{s} c_{s} = a H}
=
\left.
-
\frac{ \H^{3}}{c_{s} \P_{s} - \H^{2}}
\left(1 + \frac{\dk{c}_{s}}{c_{s}}\right)
\right|_{k_{s} c_{s} = a H}.
\label{dHLogK}
\end{gather}
We have eliminated $\epsilon$ from (\ref{DotHEqu}), but two independent variables remain. To get an equation for $c_{s}$ we need to find $\H$ in terms of the observables. Since the expression for $P_{h}$ (\ref{PhspecEquDBI}) only depends on $\H$, it can be used to find the Hubble parameter directly. In doing so, one find that
\begin{gather}
\left.
\H^{2}
\right|_{k_{t} = a H}
=
\frac{\P_{h}(k_{t})}{16}
\label{HinTermsOfPhOld}
\end{gather}
where $\P_{h} =  A^{-1} P_{h}$. This gives the Hubble parameter as a function of the tensor mode wave number $k_{t}$. In order to find $\H^{2}$ as a function of the scalar mode wave number $k_{s}$, note that  the relation between the wave number of tensor and scalar modes that exit the horizon at the same time is  $k_{t} =k_{s}  c_{s} $. Therefore, we can obtain $\left. \H \right|_{k_{s} c_{s} = a H}$ by performing the substitution $k_{t} \rightarrow k_{s} c_{s}$:
\begin{gather}
\left.
\H^{2}
\right|_{k_{s} c_{s} = a H}
=
\frac{\P_{h}(k_{s} c_{s})}{16}.
\label{HinTermsOfPh}
\end{gather}
Plugging this in for $\H$ into equation (\ref{dHLogK}) we get
\begin{gather}
\frac{d \P_{h}(k_{t})}{d \log k_{t}}
=
-
\frac{2 \P_{h}(k_{t})^{2} }{16 c_{s}(k_{s}) \P_{s}(k_{s})  - \P_{h}(k_{t})}.
\label{Phdot}
\end{gather}
Rearranging (\ref{Phdot}) we find the equation for $c_{s}$:
\begin{gather}
c_{s}(k_{s})
= 
\frac{1}{16}
\frac{\P_{h}(k_{s} c_{s})}{\P_{s}(k_{s})}
\left(
1
-
\frac{2 \P_{h}(k_{s} c_{s})}{\dkt{\P}_{h}(k_{s} c_{s})}
\right).
\label{gammaEqu}
\end{gather}
Here, a solid circle over $\P_{h}$ will denote differentiation with respect to $\log k_{t}$, not $\log k_{s}$. Once we specify what $\P_{s}(k_{s})$ and $\P_{h}(k_{t})$ are, we can use the above to solve for $c_{s}$. Even if we can only show that $c_{s} \neq 1$, this will be a signal for non-minimal kinetic terms. Once we use equations (\ref{HinTermsOfPh}) and (\ref{gammaEqu}) to get $\H(k)$ and $c_{s}(k)$ we can use (\ref{dlogkdt}) to find the relation between the wave number and time and ultimately find $\H(t)$ and $c_{s}(t)$.

Having successfully found $\H$ and $c_{s}$, the next step will be to use this information to find the action. The most general single scalar field action is a multivariable function of $\phi$ and $X$. However, since we only have $\H$ and $c_{s}$ as functions of a single independent variable (in this case $k$) the action can only be determined as a function of $k$: $p(k)$. To turn $p(k)$ into $p(\phi,X)$, we need to find $\phi$ and $X$ as functions of $k$, invert them to get $k(\phi)$ and $k(X)$, and substitute into $p(k)$. However, there is an ambiguity in how we substitute $k$ for $\phi$ and $X$. Whenever $k$ appears in the expression for $p(k)$, we do not know whether to substitute it with $k(\phi)$, $k(X)$ or some combination of the two. The ambiguity can be partially resolved if at the onset we specify a separation of the action into functions that depend either entirely on $\phi$ or entirely on $X$. In light of this fact we make an ansatz:
\begin{gather}
p(\phi , X)
=
\alpha^{2} M_{pl}^{2}
\varrho(\varphi,x)
=
\alpha^{2} M_{pl}^{2}
P(g_{1}(x),...,g_{m}(x),f_{1}(\varphi),...,f_{n}(\varphi))
\label{AssumedFormofAction}
\end{gather}
where $\varphi = M_{pl}^{-1} \phi$ and $x = (\alpha M_{pl})^{-2} X$. We have also defined a dimensionless action $\varrho$ in order to keep the exposition neat and clear. Here it is assumed that $P(y_{1},...,y_{m},z_{1},...,z_{n})$ is a known function of the $\{y_{i}\}$ and $\{z_{\alpha}\}$, and the functions $\{g_{i}\}$ and $\{f_{\alpha}\}$ are not all known. We say that (\ref{AssumedFormofAction}) only  partially resolves the ambiguity, because it is possible in certain circumstances that a field redefinition can leave the form of (\ref{AssumedFormofAction}) unalterated. For example consider the action
\begin{gather}
\varrho(\varphi,x)
=
f(\varphi)
g(x).
\end{gather}
Under a general field redefinition $\varphi = h(\tilde{\varphi})$ the action 
\begin{gather}
\varrho(\varphi,x)
=
f(h(\tilde{\varphi}))
g((h^{\prime}(\tilde{\varphi}))^{2} \tilde{x}).
\end{gather}
If the function $g$ is such that $g(x \cdot y) = g(x) \cdot g(y)$ then 
\begin{gather}
\varrho(\varphi,x)
=
f(h(\tilde{\varphi}))
g((h^{\prime}(\tilde{\varphi}))^{2}) g( \tilde{x})
=
\tilde{f}(\tilde{\varphi}) g( \tilde{x}).
\end{gather}
Thus, not all choices for the function $P$ lead to unique separation between the functions of $x$ and $\varphi$.

The first equation for the reconstructed action will be obtained from the definition of the sound speed, which is given by     
\begin{gather}
c_{s}^{2}
=
\frac{p_{,X}}{2 X p_{,XX} + p_{,X}}
\label{csInTermsOFgamma}
\end{gather}
Assuming that the action has the form (\ref{AssumedFormofAction}), equation (\ref{csInTermsOFgamma}) can now be used to find a differential equation for the $g_{i}$'s as a function of time. After some work, this equation is given by 
\begin{gather}
\left<\ddot{g}P\right>
=
\frac{\left<\dot{g}P\right> }{2}
\frac{\dot{x}}{x}
\left(
\frac{2 x \ddot{x}}{\dot{x}^{2}}
+
\frac{1}{c_{s}^{2}}
-
1
\right)
-
\left<
\dot{g} 
P 
\dot{g}
\right>
\label{ddotGEquationWithX}
\end{gather}
where a dot denotes differentation with respect to the dimensionless time $\tau = \alpha t$. Here, we have used the short hand notation:
\begin{gather}
\left<
\ddot{g} P
\right>
=
\sum_{i=1}^{m}
\ddot{g}_{i}
P_{i}
,
\qquad
\left<\dot{g}P\right>
=
\sum_{i=1}^{m}
\dot{g}_{i}
P_{i}
,
\qquad
\left<
\dot{g} P \dot{g}
\right>
=
\sum_{i,j=1}^{m}
\dot{g}_{i} 
 \dot{g}_{j}
P_{ij},
\end{gather} 
where $P_{i}$ and $P_{ij}$ are $P_{i} = \frac{\partial P}{\partial g_{i}}$ and $P_{ij} = \frac{\partial^{2} P}{\partial g_{i} \partial g_{j}}$. Equation (\ref{ddotGEquationWithX}), however, is incomplete since $x$ is not known explicitly. To turn (\ref{ddotGEquationWithX}) into a more usable form, we need to obtain an equation for $x$ and its derivatives in terms of the known quantities $\epsilon$ and $\H$. To find such a formula lets write out the expression for the energy density of a general single scalar field action (\ref{GeneralAction}):
\begin{gather}
\rho
=
2 x p_{,x} -p.
\label{GenEnergyDensity}
\end{gather}
As a consequence of the Friedmann equations,  $\rho + p$ is proportional to $\frac{d H}{dt}$:
\begin{gather}
-
\frac{d H}{d t}
=
\frac{\rho + p}{2 M_{pl}^{2}}.
\label{dHFried}
\end{gather}
Solving for $\rho + p$ in (\ref{GenEnergyDensity}), and substituting the result into (\ref{dHFried}) we find that
\begin{gather}
x
= 
-
\frac{M_{pl}^{2}}{p_{,x}}
\frac{d H}{dt}
= 
\frac{\epsilon  M_{pl}^{2} H^{2}}{p_{,x}},
\label{Xpreequation}
\end{gather}
where in the last step we have used the definition of the slow roll parameter. With (\ref{AssumedFormofAction}) as our assumed form of the action, this algebraic equation for $x$ becomes a first order differential equation:
\begin{gather}
\frac{x}{\dot{x}}
=
\frac{\epsilon \H^{2}}{\left<\dot{g}P\right>}.
\label{Xequation}
\end{gather}
Differentiating this equation, one can find an expression for $\ddot{x}$. After substituting the results of these relations, equation (\ref{ddotGEquationWithX}) becomes
\begin{gather}
\left<\dot{g}P\right>
=
\frac{2 \epsilon c_{s}^{2} \H^{2}}{\left(
1 + c_{s}^{2}
\right)}
\left(\tilde{\eta} \H
-
2 \epsilon \H
-
\frac{1}{\left<\dot{g}P\right>}
\sum_{\alpha=1}^{n}
\dot{f}_{\alpha}  \left<\dot{g}P\right>_{,\alpha}
\right),
\label{SoundSpeedEqu}
\end{gather}
where $\tilde{\eta} = \frac{\dot{\epsilon}}{\H \epsilon}$ and $\left<\dot{g}P\right>_{,\alpha}$ denotes partial differentiation of the quantity $ \left<\dot{g}P\right>$ with respect to $f_{\alpha}$. We refer to (\ref{SoundSpeedEqu}) as the {\it sound speed reconstruction equation}. The non-gaussianity parameter $f_{NL}$ can also be used to find an equation for the functions $g_{i}$ and $f_{\alpha}$. Following from the ansatz of the curvature perturbation (\ref{FnlDefinition}), $f_{NL}$ is determined by the behavior of the curvature three point function:
\begin{gather}
\left<
\zeta(\k_{1})
\zeta(\k_{2})
\zeta(\k_{3})
\right>
=
-
(2 \pi)^{7}
\delta^{(3)}(\k_{1} + \k_{2} + \k_{3})
P_{s}^{2}(K)
\frac{3 f_{NL}(K)}{10}
\frac{\sum_{i} k_{i}^{3}}{\prod_{i} k_{i}^{3}}
\label{fNLandtheThreePointFunc}
\end{gather}
where $k_{i} = |\k_{i}|$ and $K = k_{1} + k_{2} + k_{3}$. As we can see from (\ref{fNLandtheThreePointFunc}), the $f_{NL}$ will depend on the size and shape of the triangle formed by the three scales of the three point function. In \cite{Chen:2006nt} the authors found an expression for $f_{NL}$ for general single field actions (\ref {GeneralAction}) when the three scales form an equilateral triangle:
\begin{gather}
f_{NL}
=
\frac{35}{108}
\left(
\frac{1}{c_{s}^{2}}
-1
\right)
-
\frac{5}{81}
\left(
\frac{1}{c_{s}^{2}}
-1
-
2 \Lambda
\right),
\label{GeneralExpressionForfNL}
\end{gather}
where
\begin{gather}
\Lambda
=
\frac{X^{2} p_{,XX} + \frac{2}{3}X^{3} p_{,XXX}}{X p_{,X} + 2 X^{2} p_{,XX}}.
\label{ExpressionForLambda}
\end{gather}
To get $f_{NL}$ as a function of $K$, we have to evaluate (\ref{GeneralExpressionForfNL}) at the time when the scale $K$ passes outside of the sound horizon: $K c_{s} = a H$. The scalar power spectrum depends on a single scale $k$, which has a one-to-one mapping with the time through the relation $k c_{s}= a H$. However, since $f_{NL}$ really depends on three different scales, the mapping between time and scale is not as straight forward. When the delta function in (\ref{fNLandtheThreePointFunc}) is taken into account, the non-gaussianity still depends on three degrees of freedom: the magnitude of two of the scales and the angle between them \cite{Babich:2004gb}. To simplify matters, two of these three degrees of freedom will be fixed, so as to make $f_{NL}$ a univariate function. Since the equilateral configuration: $k_{1} = k_{2} = k_{3}$, has been very widely studied \cite{Chen:2006nt,Babich:2004gb}, we will take $f_{NL}$ to be the non-gaussianity of the equilateral bispectrum. The equilateral non-gaussianity will be a function of $k_{NL}$, which is the length of the sides of the equilateral triangle. Since the non-gaussianity freezes in when the scale $K$ leaves the horizon, the scales at which the non-gaussianity and the scalar perturbation freeze in are not the same but instead related by $3 k_{NL} = k_{s}$. After some work, one can use (\ref{GeneralExpressionForfNL}) and (\ref{Xequation}) to show that the $g$'s and $f$'s satisfy the equation
\begin{gather}
\left<\dot{g}P\right>
=
\frac{16 \epsilon \H^{3} c_{s}^{2}}{55}
\frac{
1}{
1
-
c_{s}^{2}
-
\frac{972}{275}
c_{s}^{2}
f_{NL}}
\nn \\
\times
\left(
 \tilde{\kappa}
+
\frac{\epsilon c_{s}^{2} \H}{ \left<
\dot{g} P \right>^{3} }
\sum_{\alpha=1}^{n}
 \dot{f}_{\alpha}  \left[\left<\dot{g}P\right> \Big(\left<\ddot{g}P\right>_{,\alpha} + \left<
\dot{g} P \dot{g}
\right>_{,\alpha} \Big) 
- 
\left<\dot{g}P\right>_{,\alpha} \Big(\left<\ddot{g}P\right>+\left<
\dot{g} P \dot{g}
\right>\Big) \right]
\right),
\label{NonGaussEqu}
\end{gather}
where $\tilde{\kappa} = \frac{\dot{c}_{s}}{\H c_{s}} = -\frac{\kappa (1 - \epsilon)}{1 - \kappa}$. This is the {\it non-gaussianity reconstruction} equation. Note that (\ref{NonGaussEqu}) is only well defined if the last line is nonzero. If the last line does vanish and the right hand side of (\ref{SoundSpeedEqu}) is nonzero then $1 - c_{s}^{2} - \frac{972}{275} c_{s}^{2} f_{NL} = 0$, and the non-gaussianity (\ref{GeneralExpressionForfNL}) only depends on the functions $g_{i}$ and $f_{\alpha}$ through the sound speed $c_{s}$. We will discuss such a case in the next section. Finally, another relation between the $f$'s and $g$'s can be derived by combining the Friedmann equations (\ref{FriedmannOne}) and (\ref{FriedmannTwo}):
\begin{gather}
P(g_{1},...,g_{m},f_{1},...,f_{n})
=
(2 \epsilon - 3) \H^{2}.
\label{FriedmannThree}
\end{gather}
The upshot is that we now have four equations: (\ref{Xequation}), (\ref{SoundSpeedEqu}), (\ref{NonGaussEqu}), and (\ref{FriedmannThree}), which when combined can be used to find $x(t)$ (and by extension $\varphi(t)$) and three of the functions $f_{\alpha}$ and $g_{i}$. With more observational inputs it may be possible to determine even more $f$ and $g$ functions, but for now we will be content with what we have. In what follows, we will consider different, specific scenarios for the action and show how the action in each can be determined from the observables.

\subsection{Examples}
\label{Examples}
In the case where the action (\ref{AssumedFormofAction}) has only one function of $x$ the equations (\ref{NonGaussEqu}) and (\ref{SoundSpeedEqu}) take on a much simpler form:
\begin{gather}
\dot{g} P_{g}
=
\frac{2 \epsilon c_{s}^{2} \H^{2}}{\left(
1 + c_{s}^{2}
\right)}
\left(\tilde{\eta} \H
-
2 \epsilon \H
-
\frac{1}{P_{g}}
\sum_{\alpha=1}^{n}
\dot{f}_{\alpha} P_{g\alpha}
\right),
\label{gdotcsEquWithOnlyOneG}
\end{gather}
\begin{gather}
\dot{g} P_{g}
=
\frac{16 \epsilon  \H^{3} c_{s}^{2}}{55}
\frac{
1}{
1
-
c_{s}^{2}
-
\frac{972}{275}
c_{s}^{2}
f_{NL}}
\left(
 \tilde{\kappa} 
+
\frac{\epsilon c_{s}^{2} \H}{ P_{g}^{3} }
\sum_{\alpha=1}^{n} \dot{f}_{\alpha}  \left[ P_{g} P_{gg\alpha} -  P_{g\alpha} P_{gg} \right]
\right).
\label{gdotfNLEquWithOnlyOneG}
\end{gather}
As we mentioned earlier not all forms of the action will yield an equation for the functions $g_{i}$ and $f_{\alpha}$. In particular if the action is such that $P_{g} P_{gg\alpha} =   P_{g\alpha} P_{gg}$ for each $\alpha$, and if the sound speed is constant, then (\ref{gdotfNLEquWithOnlyOneG}) is not well defined. To see why, lets assume that $c_{s}$ is constant. As a result, according to the definition of the sound speed:
\begin{gather}
c_{s}^{2}
=
\frac{\varrho_{,x}}{2 x \varrho_{,xx} + \varrho_{,x}} 
\quad
\Rightarrow
\quad
\varrho(\varphi,x)
=
f_{1}(\varphi) x^{\frac{1 + c_{s}^{2} }{2 c_{s}^{2}}} + f_{2}(\varphi),
\label{pWithConstantSoundSpeed}
\end{gather}
where the $f_{1}$ and $f_{2}$ are integration constants and in general will be functions of $\varphi$ only. We already know that with this form of the action $P_{g} P_{gg\alpha} =   P_{g\alpha} P_{gg}$. Thus the term in (\ref{gdotfNLEquWithOnlyOneG}) inside the large parentheses vanishes, however the right hand side of equation (\ref{gdotcsEquWithOnlyOneG}) does not vanish. Therefore, we expect the denominator $1 - c_{s}^{2} - \frac{972}{275} c_{s}^{2} f_{NL}$ in equation (\ref{gdotfNLEquWithOnlyOneG}) to vanish. Indeed if we use the formulas (\ref{GeneralExpressionForfNL}) and (\ref{ExpressionForLambda}) for $f_{NL}$ we find that
\begin{gather}
f_{NL}
=
\frac{275}{972}
\left(
\frac{1}{c_{s}^{2}}
-1
\right).
\label{IndefNonGauss}
\end{gather}
This relation between $c_{s}$ and $f_{NL}$ holds regardless of what the functions $f_{1}$ and $f_{2}$  in (\ref{pWithConstantSoundSpeed}) are. It might be argued that if $c_{s}$ is constant then the action (\ref{pWithConstantSoundSpeed}) can be assumed and the remaining equations can be used to find $f_{1}$ and $f_{2}$. This however is not that case since we have already used the sound speed equation to find $g(x)$. This can be confirmed if one assumes the action (\ref{pWithConstantSoundSpeed}). With (\ref{pWithConstantSoundSpeed}) as our action, equation (\ref{SoundSpeedEqu}) is equivalent to the time derivative of equation (\ref{Xequation}). Thus, there are really only two equations: either (\ref{Xequation}) or (\ref{SoundSpeedEqu}), and equation (\ref{FriedmannThree}). Therefore, only one of the two $f_{1}$ and $f_{2}$ can be solved for.

There is still yet another potential complication that may arise, specifically when the action takes the form $\varrho(\varphi,x) = f(\varphi) g(x)$. Using (\ref{FriedmannThree}) to find $\dot{f}$ in terms of $\dot{g}$, the equations (\ref{gdotcsEquWithOnlyOneG}) and (\ref{gdotfNLEquWithOnlyOneG}) become
\begin{gather}
\dot{g} f
=
\frac{6 \epsilon \tilde{\eta} \H^{3} c_{s}^{2}}{3 (1 + c_{s}^{2}) - 2 \epsilon},
\label{gcsEquWithGF}
\\
\dot{g} f
=
\frac{16 \epsilon \tilde{\kappa} \H^{3} c_{s}^{2}}{55} 
\frac{1}{
1
-
c_{s}^{2}
-
\frac{972}{275}c_{s}^{2} f_{NL}}.
\label{gfNLEquWithGF}
\end{gather}
As with the previous case, (\ref{gfNLEquWithGF}) is not defined when $\tilde{\kappa} = 0$. Furthermore, the first equation (\ref{gcsEquWithGF}) is also undefined when $\tilde{\eta}=0$. Since this is equivalent to $\epsilon = constant$, lets assume that $\epsilon$ is constant to see which type of action this corresponds to. From (\ref{Xpreequation})
\begin{gather}
\epsilon = \frac{3 x p_{,x}}{\rho}
\quad
\Rightarrow
\quad
\left(
2 - \frac{3}{\epsilon}
\right)
x \varrho_{,x}
=
\varrho
\quad
\Rightarrow
\quad
\varrho(\varphi,x)
=
f_{1}(\varphi)
x^{\frac{\epsilon}{2 \epsilon - 3}}.
\label{pWithConstantEpsilon}
\end{gather}
Notice, that this action is a special case of the $c_{s} = constant$ action (\ref{pWithConstantSoundSpeed}) with $f_{2} = 0$. Thus, $\epsilon = constant$ implies that $c_{s} = constant$. However, the converse of this is not true if $f_{2} \neq 0$. Since (\ref{gdotcsEquWithOnlyOneG}) is a well defined equation even with the action (\ref{pWithConstantEpsilon}), we suspect that the denominator in (\ref{gcsEquWithGF}) vanishes. After calculating the sound speed with the action (\ref{pWithConstantEpsilon}) one finds that
\begin{gather}
c_{s}^{2}
=
\frac{2 \epsilon -3 }{3}.
\end{gather}
So indeed, the denominator in (\ref{gcsEquWithGF}) does vanish. Assuming that the equations (\ref{gcsEquWithGF}) and (\ref{gfNLEquWithGF}) are well defined, consistency requires that the right hand sides of these equations be equal, leading to the relation
\begin{gather}
\frac{3 \tilde{\eta}}{3 (1 + c_{s}^{2}) - 2 \epsilon}
=
\frac{8 \tilde{\kappa}}{55} 
\frac{1}{
1
-
c_{s}^{2}
-
\frac{972}{275}c_{s}^{2} f_{NL}}.
\label{fTimesgConsistency}
\end{gather}
This is a consistency relation between $f_{NL}$, $c_{s}$ and the slow roll parameters. Although this consistency relation only holds for models with the action $\varrho = f(\varphi) g(x)$, analogous consistency relations can be found for any model in question. In what follows, we will carry out the full derivation of the $g$ and $f$ functions for two special cases.

\subsubsection{Case 1: $\varrho(\varphi,x) = g(x) - V(\varphi)$}
\label{Case1}
Suppose the action has the form 
\begin{gather}
\varrho(\varphi,x) = g(x) - V(\varphi).
\label{gVAction}
\end{gather}
This type of action corresponds to the standard scalar field action when $g$ is the identity map: $g(x) = x$. We refer to $g(x)$ as the kinetic function. Notice that we have replaced what should be $f_{1}$ in our previous nomenclature with $V(\varphi)$ in order to draw a clear analogy with the potential in the canonical scalar field action. With this type of action the equations (\ref{gdotcsEquWithOnlyOneG}) and (\ref{gdotfNLEquWithOnlyOneG}) become:
\begin{gather}
\dot{g}
=
\frac{2 \epsilon c_{s}^{2} \H^{3} \left(\tilde{\eta}
-
2 \epsilon
\right)}{1 + c_{s}^{2}},
\label{gMinusVcsEquation}
\\
\dot{g}
=
\frac{16 \epsilon  \tilde{\kappa}  \H^{3} c_{s}^{2}}{55}
\frac{
1}{
1
-
c_{s}^{2}
-
\frac{972}{275}
c_{s}^{2}
f_{NL}}.
\label{SingledotGfNLEquation}
\end{gather}
Here we have two expressions for the derivative of $g(\tau)$. Consistency demands that the right hand sides of these equations be equal, thus we are lead to an analogue of the relation (\ref{fTimesgConsistency}):
\begin{gather}
\frac{\tilde{\eta}
-
2 \epsilon}{1 + c_{s}^{2}}
=
\frac{8  \tilde{\kappa}}{55}
\frac{
1}{
1
-
c_{s}^{2}
-
\frac{972}{275}
c_{s}^{2}
f_{NL}}.
\label{gVRelation}
\end{gather} 
Interestingly enough this is the same consistency relation found in \cite{Bean:2008ga} for general single field inflation models. However, while the relation in \cite{Bean:2008ga} was only approximate, in our case it is exact. This shows that the consistency relation of Bean {\it et al.} is exact in the case when the action is of the form (\ref{gVAction}). Continuing with the derivation, the equations for $V(\tau)$ and $x(\tau)$ are given by (\ref{FriedmannThree}) and (\ref{Xequation}), respectively. They read
\begin{gather}
V(\tau) = g(\tau) - (2 \epsilon - 3) \H^{2},
\label{SingleGFriedmannEqu}
\\
\frac{\dot{x}}{\H x}
=
\frac{\dot{g}}{\epsilon \H^{3}}
=
\frac{2 c_{s}^{2} (\tilde{\eta} - 2 \epsilon)}{1 + c_{s}^{2}}.
\end{gather}
With our equations in hand we are almost ready to solve them and find the action. However, we still lack knowledge about the $\H$ and $c_{s}$. In order to go further we need to look back to section \ref{EqusofInflation} and in particular equations (\ref{HinTermsOfPh}) and (\ref{gammaEqu}). In order for these equations to be of any use we need two observables as inputs. For these we will assume that the two inputs are the scalar and tensor contributions to the CMBR. Presently, it is believed that these spectra are near scale invariant, and over a limited range of scales possess the forms\footnote{Note that the scale invariant tensor spectrum is most commonly defined at $n_{T} = 0$, whereas we have defined it at $n_{T} = 1$. The reason for using a nonstandard definition of $n_{T}$ is to establish a parity between the forms of the scalar and tensor perturbation spectra. The relation between the two different $n_{T}$'s is simply $\left.n_{T}\right|_{standard} = \left.n_{T}\right|_{ours} - 1$.}
\begin{gather}
\P_{s}(k)
=
e^{(n_{s} - 1) \log k/k_{0}},
\label{PscalarApproxOne}
\\
\P_{h}(k)
=
B e^{(n_{T} - 1) \log k c_{s}(k) / k_{0} c_{s 0}}.
\label{PtensorApproxOne}
\end{gather}
Here $c_{s 0}  = c_{s}(k_{0})$ where $k_{0}$ is the fiducial scale at which $\P_{s} = 1$. Note that we have assumed that the spectral indices have no running: i.e. $n_{s}, n_{T} = constant$. Although running spectral indices is an interesting generalization, in order to better demonstrate the usefulness of this procedure we will stick with simpler case of no running. Equation (\ref{HinTermsOfPh}) then tells us that the Hubble parameter is simply proportional to the square root of the tensor perturbation: 
\begin{gather}
\H(k)
=
\frac{\sqrt{\P_{h}(k) }}{4}
=
\frac{\sqrt{B}}{4}
e^{\frac{n_{T} - 1}{2}\log k c_{s}(k) / k_{0} c_{s 0}  }.
\label{FirstHinTermsofCs}
\end{gather}
Defining $\H_{0}$ as $\H(k_{0}) = \H_{0}$, the constant $B$ is therefore $B = 16 \H_{0}^{2}$. As it stands, (\ref{FirstHinTermsofCs}) is not complete since we still do not have an expression for $c_{s}(k)$. To find $c_{s}$ we turn to equation (\ref{gammaEqu}), the solution of which gives us
\begin{gather}
c_{s}
=
c_{s 0}
\left(
\frac{c_{s 0}}{\H_{0}^{2}}
\frac{n_{T} - 1}{ n_{T} - 3}
\right)^{\frac{1}{n_{T} - 2}}
e^{-\frac{n_{T} - n_{s}}{n_{T} - 2} \log k /k_{0}}.
\end{gather}
Since we defined $c_{s0}$ as $c_{s}(k_{0}) = c_{s 0}$, consistency of our definition demands that
\begin{gather}
c_{s 0}
=
\H_{0}^{2}
\frac{ n_{T} - 3}{n_{T} - 1}.
\end{gather}
Note that if $1 < n_{T} < 3$, $c_{s 0}$ is negative: a nonsense result. Therefore, we must restrict $n_{T}$ to be either $n_{T} < 1$ or $n_{T} >3$. With an expression for $c_{s}$ in hand, $\H(k)$ explicitly in terms of $k$ is
\begin{gather}
\H(k)
=
\H_{0}
e^{\frac{(n_{T} - 1)}{2} \frac{n_{s} - 2}{n_{T} - 2} \log k /k_{0} }.
\end{gather}
Note that $\epsilon$ and $\kappa$ are constant in this case and are given by
\begin{gather}
\epsilon
=
\frac{\H^{2}}{c_{s} \P_{s}}
=
\frac{n_{T} - 1}{n_{T} - 3},
\label{Epsilon}
\\
\kappa 
= 
-
\frac{\dk{c}_{s}}{c_{s}}
=
\frac{n_{T} - n_{s}}{n_{T} - 2}.
\end{gather}
Since $\epsilon$ is a constant then $\tilde{\eta} = 0$, which will simplify matters later when we try solve the reconstruction equations. Solving for $\log k$ in (\ref{dlogkdt}), we find that the time dependence of $\log k$ is
\begin{gather}
\log k/k_{0}
=
\frac{\omega}{\kappa}
\log\left[
1
+
\epsilon
\H_{0}
(\tau - \tau_{0})
\right],
\label{logkInTermsofTime}
\end{gather}
where $\omega = \frac{\kappa}{\epsilon} \frac{1 - \epsilon}{1 - \kappa}= - \frac{2 (n_{T} - n_{s})}{(n_{T} - 1) (n_{s} - 2)}$. Therefore, the sound speed and Hubble parameter as functions of time are
\begin{gather}
c_{s}(\tau)
=
c_{s0}
\left[
1
+
\epsilon
\H_{0}
(\tau - \tau_{0})
\right]^{- \omega}
,
\qquad
\H(\tau)
=
\frac{\H_{0}}{
1
+
\epsilon
\H_{0}
(\tau - \tau_{0})}.
\label{CsandHinTermsofTime}
\end{gather}
These are the expressions for the sound speed and Hubble parameter that will be used throughout this paper. They are completely independent of the form of the action that we are solving for, and are determined only by the inflationary observables $\P_{s}$ and $\P_{h}$.

Before we go about solving the reconstruction equations we should point out that not all values of the spectral indices lead to realistic inflationary scenarios. As has been mention before, the sound horizon of the scalar fluctuations is not the same as the cosmological horizon. As a consequence it is now possible for the size of the sound horizon to increase as time progresses. Thus, the usual expectation that larger scales freeze in at the beginning of inflation and smaller scales freeze in at the end, is not always guaranteed to hold. Recall that the time dependence of the scale is given by equation (\ref{logkInTermsofTime}). It follows that the sound horizon size depends on time like
\begin{gather}
\textrm{Sound Horizon Size}
\propto
(\gamma a H)^{-1}
=
(k/k_{0})^{-1}
=
\left(
1
+
\epsilon
\H_{0}
(\tau - \tau_{0})
\right)^{-\frac{\omega}{\kappa}}.
\end{gather}
If $\omega / \kappa >0$, the sound horizon decreases with time as is normally expected. However, if $\omega /\kappa < 0$, the size of the sound horizon increases during inflation, allowing modes the were previously frozen-in behind the horizon to reenter while inflation is still going on. This would be a disaster since if the scales were to reenter during inflation they would continue to fluctuate, destroying the near-scale invariance of the CMB anisotropy. Therefore, the spectral indices must be fixed such that $\omega$ and $\kappa$ are either both positive or both negative.

Since we are considering only those models that allow for inflation, we need to be sure that the spectral indices are such that an inflationary phase is allowed. If we refer to the expression for the equation of state $w$ we see that not all values of $n_{T}$ are allowed if we want to have inflation:
\begin{gather}
w
=
\frac{p}{\rho}
=
\frac{p}{2 X p_{,X} - p}
=
-
\frac{n_{T} - 7}{3 (n_{T} - 3)}.
\end{gather}
Notice that so long as $n_{T} < 3$ the equation of state is always $w < - \frac{1}{3}$, and so inflation will occur. Since $c_{s} \propto \epsilon^{-1}$, in order for $c_{s}$ to be interpreted as a sound speed, $\epsilon$ must be positive. If we look back to equation (\ref{Epsilon}) we find that not all values of $n_{T}$ will result in a positive value for $\epsilon$. Requiring that $\epsilon>0$, we find that $n_{T}$ must be either $n_{T} < 1$ or $n_{T} >3$. Since we have already found that $n_{T} >3$ would not lead to an inflationary solution, we conclude that $n_{T}  < 1$. Recall that in the previous paragraph we found that the sound horizon could expand during inflation only if the spectral indices were chosen so that $\omega / \kappa > 0$. If one refers back to the definitions of $\omega$ and $\kappa$ in terms of the spectral indices, we can see that if $n_{T}  < 1$ the scalar spectral index is required to be $n_{s} <2$.

The sound speed (\ref{CsandHinTermsofTime}) can tell us something about the expected range of validity of the scalar (\ref{PscalarApproxOne}) and tensor spectra (\ref{PtensorApproxOne}). If $\omega < 0$  then at some time $\tau >\tau_{0}$ the sound speed will be greater than one, signaling that fluctuations propagate at superluminal speeds. Likewise, superluminal speeds also occur at times $\tau < \tau_{0}$ when $\omega >0$. Therefore,  (\ref{PscalarApproxOne}) and (\ref{PtensorApproxOne}) can only be considered approximations; reliable within a certain range of scales. Keeping in mind that the sound horizon needs to shrink during inflation, the wave number $k$ must respect the following bounds if the sound speed is to be less then the speed of light: 
\begin{align}
k/k_{0}
>
(c_{s0})^{\frac{1}{|\kappa|}},
\qquad
&\textrm{for  }\, \omega >0,
\nn \\
k/k_{0}
<
(c_{s0})^{-\frac{1}{|\kappa|}},
\qquad
&\textrm{for  }\, \omega < 0.
\label{CsKbounds}
\end{align}
The only way (\ref{PscalarApproxOne}) and (\ref{PtensorApproxOne}) could be acceptable at all scales is if $\omega =0$, in which case $c_{s}$ is a constant. If it turns out that the sound speed is not constant, (\ref{PscalarApproxOne}) and (\ref{PtensorApproxOne}) are most likely too naive. The most recent data from WMAP \cite{Komatsu:2008hk} suggests that the scalar spectral index may have a small but nonzero running, so we should not be surprised that our simple expressions for the perturbation spectra are not exactly correct. Regardless, scalar and tensor spectra with constant spectral indices are still a good approximation to the CMBR data. Our discussion will still be of relevance, as long as we keep in mind that the reconstructed actions are only approximations, valid over a limited range of scales.

We will now simplify our discussion by fixing the sound speed to a constant, which is achieved by setting $\omega = 0$. Although we will be considering only constant sound speeds, we will keep the value of $c_{s}$ arbitrary. This will allow us to find a more general solution to the reconstruction equations, while allowing us to study the limit $c_{s} \rightarrow 1$ and see whether the canonical action is recovered. One might object to this choice of $\omega$ on the grounds that if $c_{s}$ is constant the non-gaussianity reconstruction equation (\ref{SingledotGfNLEquation}) will be ill-defined for reasons discussed in section \ref{Examples}. However, we counter that this is acceptable since we are assuming that only two functions $g$ and $V$ are unknown, thereby making the third reconstruction equation (\ref{SingledotGfNLEquation}) unnecessary. It should be pointed out that while the two unknown functions can still be found when $\omega = 0$, the consistency relation (\ref{gVRelation}) is no longer well defined. Once we substitute (\ref{CsandHinTermsofTime}) for $c_{s}$ and $\H$ into (\ref{gMinusVcsEquation}) and solve for $g(\tau)$ the result is
\begin{gather}
g(\tau)
=
\frac{2 \epsilon \H_{0}^{2} c_{s0}^{2}}{1 + c_{s0}^{2}}
\left(
\frac{1}{(1 + \epsilon \H_{0} (\tau - \tau_{0}))^{2}}
-
1
\right)
+
g_{0}.
\label{gtau}
\end{gather}
Using the Friedmann equation (\ref{SingleGFriedmannEqu}) we can now find $V(\tau)$:
\begin{gather}
V(\tau)
=
\frac{\H_{0}^{2}}{1 + c_{s0}^{2}}
\left(
\frac{3 (1 + c_{s0}^{2}) - 2 \epsilon}{(1 + \epsilon \H_{0} (\tau - \tau_{0}))^{2}}
-
2 \epsilon c_{s0}^{2}
\right)
+
g_{0}.
\label{Vtau}
\end{gather}
Similarly, the equation for $\dot{x}$ is given by
\begin{gather}
\frac{\dot{x}}{x}
=
-
\frac{4 \epsilon c_{s0}^{2}}{1 + c_{s0}^{2}}
\frac{\H_{0}}{
[1
+
\epsilon
\H_{0}
(\tau - \tau_{0})]},
\end{gather}
and the exact solution for $x$ is
\begin{gather}
x(\tau)
=
x_{0}
\left[
1 + \epsilon \H (\tau - \tau_{0})
\right]^{- \frac{4 c_{s0}^{2}}{1 + c_{s0}^{2}}}.
\label{XwithOmegaEqZero}
\end{gather}
Integrating (\ref{XwithOmegaEqZero}) to find $\varphi(\tau)$
\begin{gather}
\varphi
=
\frac{\dot{\varphi}_{0}}{\epsilon \H_{0}}
\frac{1 + c_{s0}^{2}}{1 - c_{s0}^{2}}
\left[
(1 + \epsilon \H_{0} (\tau - \tau_{0}))^{\frac{1 - c_{s0}^{2}}{1 + c_{s0}^{2}}}
-
1
\right]
+
\varphi_{0},
\end{gather}
where $\dot{\varphi}_{0} = \sqrt{2 x_{0}}$. Now that we have $x$ and $\varphi$ as functions of time, we can invert these and substitute the results into (\ref{gtau}) and (\ref{Vtau}) to find $g(x)$ and $V(\varphi)$. After carrying this out, we can combine the $g(x)$ and $V(\varphi)$ to arrive at the full action:
\begin{gather}
\varrho(\varphi,x)
=
\frac{2 \epsilon \H_{0}^{2} c_{s0}^{2}}{1 + c_{s0}^{2}}
(x/x_{0})^{\frac{1 + c_{s0}^{2}}{2 c_{s0}^{2}} }
-
\frac{\H_{0}^{2} [3 (1 + c_{s0}^{2}) - 2 \epsilon] }{1 + c_{s0}^{2}}
\left[
1
+
\epsilon \H_{0}
\frac{1 - c_{s0}^{2}}{1 + c_{s0}^{2}}
\frac{\varphi - \varphi_{0}}{\dot{\varphi}_{0}}
\right]^{- \frac{2 (1 + c_{s0}^{2})}{1 - c_{s0}^{2}}}.
\end{gather}
Here is the complete action in the case when the sound speed is constant. Note that the final result does not depend on the integration constant $g_{0}$. This is a result of the fact that the right hand side of equation (\ref{FriedmannThree}) is independent of the initial values of the kinetic and potential functions. The only undetermined constants are the initial values of the scalar field and it's derivative, and due to the attractor nature of inflation, their exact values are unimportant. Despite what was said earlier in regards to the indefiniteness of the non-gaussianity reconstruction equation (\ref{SingledotGfNLEquation}), this action does have a definite non-gaussianity given by the result in equation (\ref{IndefNonGauss}). It is worth noting that in the exceptional case where $c_{s}\rightarrow 1$:
\begin{gather}
\varrho(\varphi,x)
=
\epsilon \H_{0}^{2}
(x/x_{0})
-
\H_{0}^{2} (3 - \epsilon)
e^{- 
\frac{2
\epsilon \H_{0}}{\dot{\varphi}_{0}}
(\varphi - \varphi_{0})
}
\end{gather}
we recover the canonical inflation action with an exponential potential. If we require that the wave function retain the standard normalization then $x_{0} = \epsilon \H_{0}^{2}$ and the action becomes
\begin{gather}
\varrho(\varphi,x)
=
x
-
\H_{0}^{2} (3 - \epsilon)
e^{- 
\sqrt{2
\epsilon}
(\varphi - \varphi_{0})
},
\label{ExpCanonicalAction}
\end{gather}
which is the action of power law inflation \cite{Lucchin:1984yf}. This is a reassuring result; it confirms that in the appropriate limit, we can recover the standard inflationary action.

\subsubsection{Case 2: $\varrho(\varphi,x) = f(\varphi) g(x) - V(\varphi)$}
Let us now take the complexity of the action one step further and assume that there are now three unknown functions: $g$, $f$ and $V$. We define $\varrho$ as
\begin{gather}
\varrho(\varphi,x) = f(\varphi) g(x) - V(\varphi).
\label{1FormofAction}
\end{gather}
In equation (\ref{gdotcsEquWithOnlyOneG}) the only nonzero $P_{g\alpha}$ is the one corresponding to $f$. Thus (\ref{gdotcsEquWithOnlyOneG}) reduces to
\begin{gather}
\dot{g} f
=
\frac{2 \epsilon c_{s}^{2} \H^{2}}{\left(
1 + c_{s}^{2}
\right)}
\left(\tilde{\eta} \H
-
2 \epsilon \H
-
\frac{\dot{f}}{f}
\right)
\label{gdotcsEquWithGfV}
\end{gather}
Furthermore, with this action the terms with the $\dot{f}_{\alpha}$'s in (\ref{gdotfNLEquWithOnlyOneG}) all vanish. The final result is simply:
\begin{gather}
\dot{g} f
=
\frac{16 \epsilon  \tilde{\kappa}  \H^{3} c_{s}^{2}}{55}
\frac{
1}{
1
-
c_{s}^{2}
-
\frac{972}{275}
c_{s}^{2}
f_{NL}}.
\label{gdotfNLEquWithGfV}
\end{gather}
Combining equations (\ref{gdotcsEquWithGfV}) and (\ref{gdotfNLEquWithGfV}), $g$ decouples and we get equation just for $f$:
\begin{gather}
\frac{\dot{f}}{\H f}
=
\tilde{\eta}
-
2 \epsilon
-
\frac{8  \tilde{\kappa}}{55}
\frac{
1 + c_{s}^{2}}{
1
-
c_{s}^{2}
-
\frac{972}{275}
c_{s}^{2}
f_{NL}}.
\label{fdotEquationFromfNL}
\end{gather}
Once we have solved for $f$ here we can substitute the solution into equation (\ref{gdotfNLEquWithGfV}) and solve for $g$. With the solutions for these two, $V$ is found using the Friedmann equation (\ref{FriedmannThree}). The final step is to find $\varphi(\tau)$ and $x(\tau)$ by solving (\ref{Xequation}):
\begin{gather}
\dot{x}
=
\frac{16 \tilde{\kappa} \H c_{s}^{2}}{55}
\frac{x}{1 - c_{s}^{2} - \frac{972}{275} c_{s}^{2} f_{NL}}.
\label{XdotEquationFromfNL}
\end{gather}
Let's assume that $\tilde{\kappa} \neq 0$, so that the reconstruction equations (\ref{gdotfNLEquWithGfV}) (\ref{fdotEquationFromfNL}) and (\ref{XdotEquationFromfNL}) are well defined. We will again assume that the scalar and tensor perturbation spectra are given by (\ref{PscalarApproxOne}) and (\ref{PtensorApproxOne}). Therefore, $\H$ and $c_{s}$ are the same as those that we found earlier (\ref{CsandHinTermsofTime}). However, now that we are using the non-gaussianity reconstruction equation we need to specify $f_{NL}$. In this example we will take $f_{NL}=0$ to simplify the analysis. With these as our inputs, the reconstruction equations become
\begin{gather}
\frac{\dot{f}}{\H f}
=
-
2 \epsilon
+
\frac{8  \epsilon \omega}{55}
\frac{
1 + c_{s}^{2}}{
1
-
c_{s}^{2}}
,
\qquad
\dot{g} f
=
-
\frac{16 \epsilon^{2} \omega}{55}
\frac{ c_{s}^{2} \H^{3}}{
1
-
c_{s}^{2}}
,
\qquad
\frac{\dot{x}}{\H x}
=
-
\frac{16 \epsilon \omega}{55}
\frac{c_{s}^{2}}{1 - c_{s}^{2}}.
\label{fgxEquationsForGtimesFminusV}
\end{gather}
Here, we have used the fact that $\tilde{\eta} = 0$ and $\tilde{\kappa} = - \epsilon \omega$. Each of these has an analytic solution. They are
\begin{gather}
f(\tau)
= 
f_{0}
\left[ \frac{c_{s}^{2}(\tau)}{c_{s0}^{2}} \right]^{\frac{1}{\omega} - \frac{4}{55}}
\left(
\frac{ 1 - c_{s}^{2}(\tau) }{ 1 - c_{s0}^{2} } \right)^{8/55},
\\
g(\tau)
=
g_{0}
+
\frac{2  \H_{0}^{4} (1 - c_{s0}^{- 2})^{\frac{8}{55}}}{c_{s0} f_{0}}
\left[
\left[ \frac{c_{s}^{2}(\tau)}{c_{s0}^{2}} \right]^{- \frac{4}{55}}
F(c_{s}^{-2}(\tau))
-
F(c_{s0}^{-2})
\right],
\label{gGtimesFminusVSolution}
\\
x(\tau)
=
x_{0}
\left(
\frac{ 1 - c_{s0}^{2} }{ 1 - c_{s}^{2}(\tau) }
\right)^{\frac{8}{55}}.
\label{xGtimesFminusVSolution}
\end{gather}
Here we have defined $F(x)$ as
\begin{gather}
F(x)
=
{}_{2}F_{1}(\frac{4}{55},\frac{63}{55},\frac{59}{55}; x)
\end{gather}
where ${}_{2}F_{1}$ is a hypergeometric function. Note that we can find a complete expression for $g(x)$. We simply have to solve for $c_{s}(\tau)$ in (\ref{xGtimesFminusVSolution}) to get $c_{s}^{2}(x)$, which is 
\begin{gather}
c_{s}^{2}(x)
=
1 - (1 - c_{s0}^{2}) \left(\frac{x}{x_{0}}\right)^{-55/8},
\label{CsInTermsofX}
\end{gather}
and then substitute this for $c_{s}(\tau)$ in (\ref{gGtimesFminusVSolution}) to get $g(x)$. Interestingly enough, $g(x)$ is independent of $\omega$, so taking the $\omega \rightarrow 0$ limit here is trivial. Since the solution for $g(x)$ is in terms of hypergeometric functions, to get a better idea of what $g(x)$ looks like we expand around $c_{s0} = 1$, and thus obtain  
\begin{gather}
g(x)
=
g_{0}
+
\frac{\epsilon \H_{0}^{2}}{f_{0} c_{s0}}
\frac{x - x_{0}}{x_{0}}
+
\frac{\epsilon \H_{0}^{2} (1 - c_{s0})}{2585 f_{0} c_{s0}}
\left[
2145
-
2209
\frac{x}{x_{0}}
+
64
\left(
\frac{x_{0}}{x}
\right)^{\frac{47}{8}}
\right]
+
\O((1 - c_{s0})^{2}).
\label{gGtimesFminusVApprox}
\end{gather}
Let's take a moment to comment on the analytic behavior of $g(x)$. 
\begin{figure}
\begin{center}
\includegraphics[scale=.65]{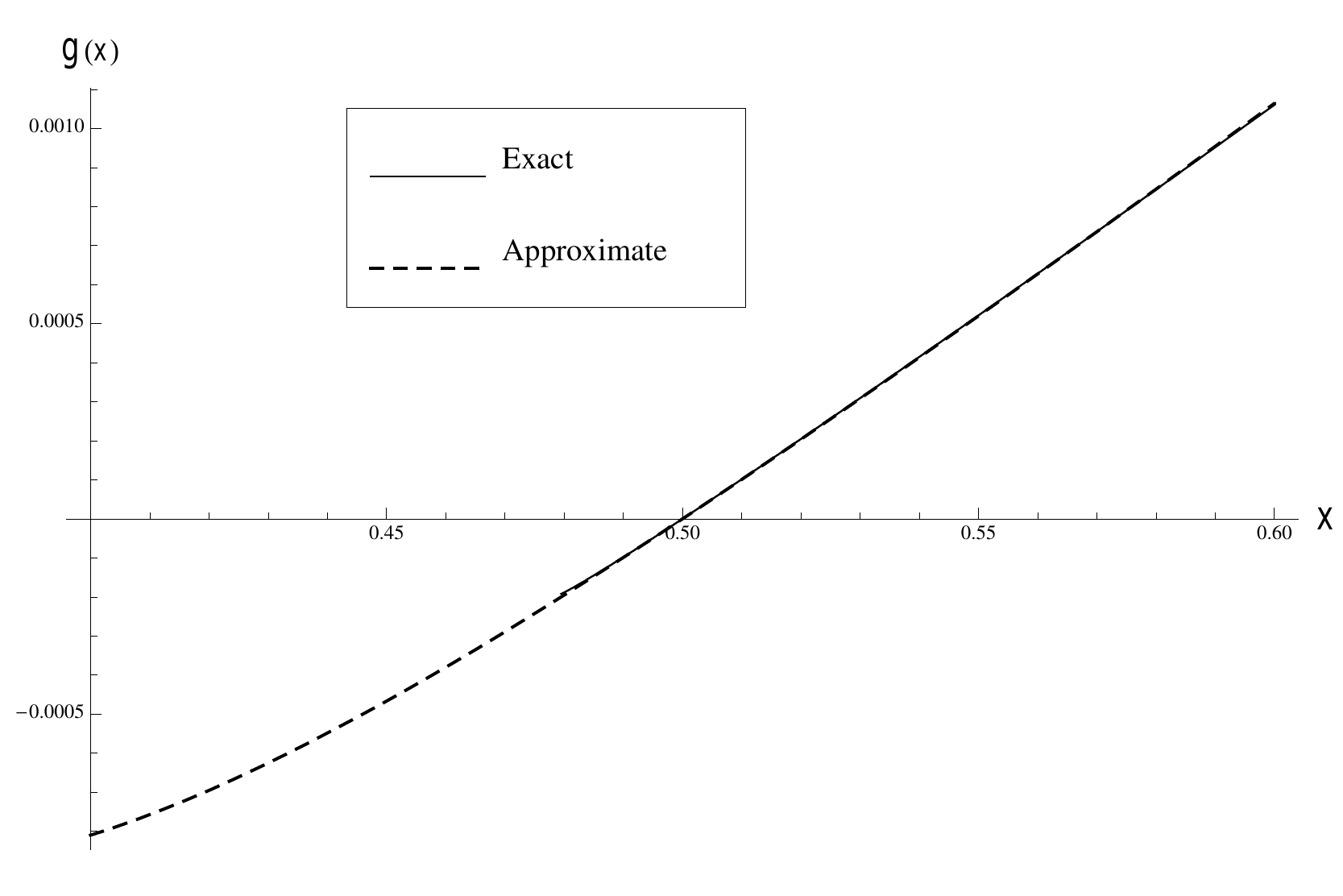}
\caption{Plot depicting the function $g(x)$. This plot was made with $c_{s0} = 0.5$, $\epsilon = 0.1$, $f_{0} = 1$, $g_{0}=0$ and $\dot{\varphi}_{0} =1$. The exact behavior of $g(x)$ is contrasted against the approximation (\ref{gGtimesFminusVApprox}). The behavior of $g(x)$ is very linear except for small deviations for $x < x_{0}$. Note that at $x_{0} (1 - c_{s0}^{2})^{8/55} \approx 0.48$ the plot of the exact behavior of $g(x)$ stops abruptly as a result of the fact that $g$ becomes non-real in this region. }
\label{GVersusXFTimeGMinusVAction}
\end{center}
\end{figure}
In fig. \ref{GVersusXFTimeGMinusVAction} the exact functional behavior of $g(x)$ is shown along with the approximate expression (\ref{gGtimesFminusVApprox}). As fig. \ref{GVersusXFTimeGMinusVAction}  and the approximation (\ref{gGtimesFminusVApprox}) suggest, the behavior of $g$ is nearly linear with respect to $x$, when $c_{s0}$ is close to one. However, for reasons that will be clear shortly, the limit $c_{s0} \rightarrow 1$ does not necessarily mean that the {\it action} will be linear in $x$. Another interesting feature of $g(x)$ is that it becomes non-real for values of $x$ less than $x_{0} (1 - c_{s0}^{2})^{8/55}$. This implies a lower bound on the values of $x$, which is a behavior that is observed in the solution (\ref{xGtimesFminusVSolution}). This lower bound is a result of the fact that for $x < x_{0} (1 - c_{s0}^{2})^{8/55}$ the sound speed squared would be negative according to (\ref{CsInTermsofX}).

As for the functions $f(\varphi)$ and $V(\varphi)$, one cannot find analytic expressions for these like we did for $g(x)$. Once we integrate $x(\tau)$ to find $\varphi(\tau)$, we can see why:
\begin{gather}
\varphi
=
\varphi_{0}
+
\frac{\sqrt{2x_{0}} \, c_{s0} (1 - c_{s0}^{- 2})^{4/55}}{\H_{0}^{3} (1 + \frac{8 \omega}{55})}
\left[
\left[\frac{c_{s}^{2}(\tau)}{c_{s0}^{2}} \right]^{-\frac{1}{2 \omega} - \frac{4}{55}}
F_{\omega}(c_{s}^{-2}(\tau) )
-
F_{\omega}(c_{s0}^{-2})
\right],
\label{phiGtimesFminusVSolution}
\end{gather}
where we have again shortened things by defining $F_{\omega}$ as
\begin{gather}
F_{\omega}(x)
=
{}_{2}F_{1}(\frac{4}{55} + \frac{1}{2 \omega}, \frac{4}{55}, \frac{59}{55} + \frac{1}{2 \omega}; x).
\end{gather}
Since $\varphi$ is such a complicated function there is no way to invert (\ref{phiGtimesFminusVSolution}) to get time as an analytic function of $\varphi$. Therefore, we are forced to either evaluate $f(\varphi)$ and $V(\varphi)$ numerically, or make an approximation for $\tau(\varphi)$. Since we will be interested in finding a correspondence with the example in the previous section, we will approximate $\varphi(\tau)$ in the $\omega \ll 1$ limit. The result of this approximation is
\begin{gather}
\phi(\tau)
=
\phi_{0}
+
\dot{\phi}_{0} \tau
-
\frac{8 \dot{\phi}_{0} \omega }{55 \epsilon \H_{0} }
\frac{c_{s0}^{2}}{1 - c_{s0}^{2}}
\left[
(1 + \epsilon \H_{0} \tau)
\log (1 + \epsilon \H_{0} \tau)
-
\epsilon \H_{0} \tau
\right]
+
\mathcal{O}(\omega^{2}).
\label{PhiOmegaApproximation}
\end{gather}
To get $\tau(\varphi)$ we will drop all $\omega$ dependent terms from (\ref{PhiOmegaApproximation}), so that $\varphi(\tau)$ is a linear function of $\tau$. This approximation turns out to be remarkably accurate even at late times, since the higher order terms in (\ref{PhiOmegaApproximation}) scale only logarithmically with $\tau$. Now that we have at least an approximate expression for $\tau(\varphi)$, $f(\varphi)$ can be found by replacing $c_{s}(\tau)$ with
\begin{gather}
c_{s}(\varphi)
=
c_{s0}
\left(
1 + \frac{\epsilon \H_{0}}{\dot{\varphi}_{0}} (\varphi - \varphi_{0})
\right)^{- \omega}.
\label{csPhiGtimesFminusV}
\end{gather}
The exact behavior of $f(\varphi)$ was evaluated numerically and the results are shown in 
\begin{figure}
\begin{center}
\includegraphics[scale=.65]{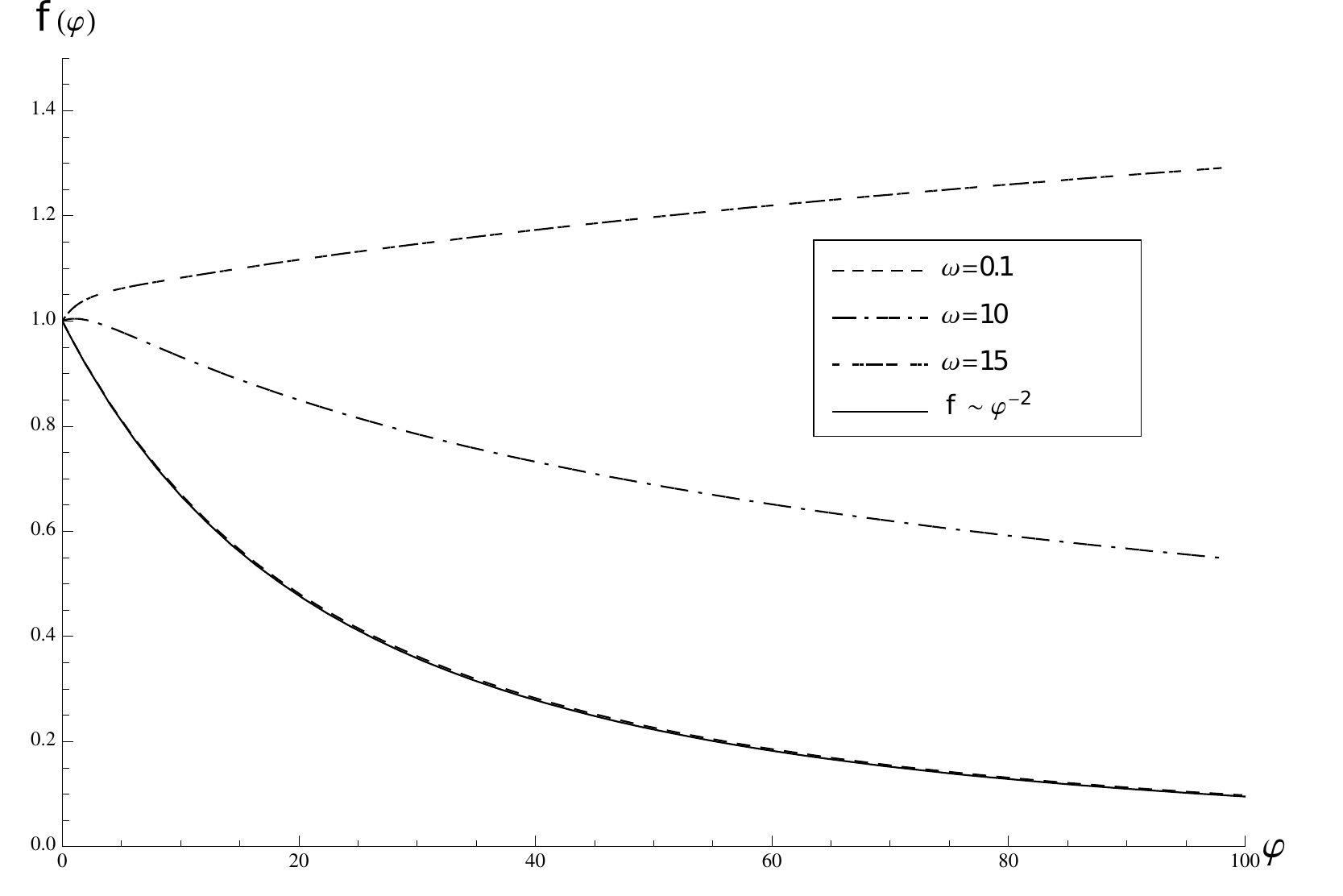}
\caption{Plot depicting the function $f(\varphi)$ for different values of $\omega$. This plot was made with $c_{s0} = 0.5$, $\epsilon = 0.1$, $\varphi_{0} = 0$, $\dot{\varphi}_{0} = 1.0$, $f_{0} = 1.0$ and $g_{0} = 0$. We have also included a plot of the function $f(\varphi)$ in (\ref{gfVinCanonicalLim}) for comparison.}
\label{FVersusPhiForFtimesGMinusV}
\end{center}
\end{figure}
fig. \ref{FVersusPhiForFtimesGMinusV}. Since we will be taking the $\omega \rightarrow 0$ limit later, we make a further approximation of $f(\varphi)$ by taylor expanding around $\omega = 0$:
\begin{gather}
f(\varphi)
\approx
\frac{f_{0}}{(1 + \frac{\epsilon \H_{0}}{\dot{\varphi}_{0}} (\varphi - \varphi_{0}))^{2}}
+
\frac{8 \omega f_{0}}{55}
\frac{1 + c_{s0}^{2}}{1 - c_{s0}^{2}}
\frac{\log [1 + \frac{\epsilon \H_{0}}{\dot{\varphi}_{0}} (\varphi - \varphi_{0})]}{(1 + \frac{\epsilon \H_{0}}{\dot{\varphi}_{0}} (\varphi - \varphi_{0}))^{2}}.
\label{fGtimesFminusVApprox}
\end{gather}
Notice that the second term diverges at $c_{s0} = 1$. Therefore, although the higher order terms in (\ref{gGtimesFminusVApprox}) vanish when $c_{s0}=1$, when the limit $c_{s0} \rightarrow 1$ is taken the product $f g$ will retain the nonlinear $x$ terms. This is why the action may not be linear in $x$ even when $c_{s0} = 1$. To find the potential we use the Friedmann equation (\ref{FriedmannThree}). Doing so requires us to find $g$ as a function of $\varphi$, which we find by replacing $c_{s}(\tau)$ in (\ref{gGtimesFminusVSolution}) with $c_{s}(\varphi)$ (\ref{csPhiGtimesFminusV}). The potential, taylor expanded around $\omega = 0$, is
\begin{gather}
V(\varphi)
\approx
\frac{ g_{0} f_{0} - ( 2 \epsilon - 3) \H_{0}^{2}}{(1 + \frac{\epsilon \H_{0}}{\dot{\varphi}_{0}} (\varphi - \varphi_{0}))^{2}}
+
\frac{8 \omega}{55}
\frac{ g_{0} f_{0}(1 + c_{s0}^{2}) - 2 \epsilon \H_{0}^{2} c_{s0}}{1 - c_{s0}^{2}}
\frac{ \log[1 + \frac{\epsilon \H_{0}}{\dot{\varphi}_{0}} (\varphi - \varphi_{0})]}{(1 + \frac{\epsilon \H_{0}}{\dot{\varphi}_{0}} (\varphi - \varphi_{0}))^{2}}.
\label{VGtimesFminusVApprox}
\end{gather}
Note that the individual functions $g$, $f$ and $V$ depend on the arbitrary integration constants $f_{0}$ and $g_{0}$, even though the action that is composed of them does not. If we are interested in just finding the action, fixing $f_{0}$ and $g_{0}$ would be a moot point. However, it does raise the matter of how one separates the action into kinetic and potential terms. For example, suppose we separate the kinetic function into a constant and a ``$x$-dependent'' piece:
\begin{gather}
g(x)
=
c
+
G(x).
\label{gSep}
\end{gather}
The constant $c$ is arbitrary and can be adjusted to any given value by absorbing the difference into $G(x)$. Substituting the right hand side of (\ref{gSep}) for $g(x)$, the action (\ref{1FormofAction}) becomes 
\begin{gather}
\varrho(\varphi,x)
=
f(\varphi) G(x)
-
V(\varphi)
+
c
f(\varphi).
\end{gather}
With the action written in this way, it would make more sense to define $G(x)$ as the kinetic function and define the potential as 
\begin{gather}
v(\varphi)
=
V(\varphi)
-
c
f(\varphi).
\label{RedefPotential}
\end{gather}
In the case where $f$ is constant (such as the example in the previous section) then the above redefinition only amounts to a uniform shift in the potential. However, if $f$ is non-constant then the behavior of the potential can change drastically. Although none of the CMBR data are sensitive to changes in $c$, it is possible to find a value for $c$ by requiring that in the appropriate limit, the action becomes equivalent to the canonical action. We will define this as the {\it canonical limit} of the action. Before we determine $c$ by this method we need to confirm that the action (\ref{1FormofAction}) is canonically equivalent to the canonical action (\ref{ExpCanonicalAction}) when the sound speed is constant and equal to one.

If we turn our attention back to our approximations for $f(\varphi)$ and $V(\varphi)$, we notice that taking $c_{s0} = 1$ leads to divergent results. These divergence are understandable since the reconstruction equations (\ref{fgxEquationsForGtimesFminusV}) are divergent when $c_{s} = 1$. However, if we set $\omega = 0$ in (\ref{fGtimesFminusVApprox}) and (\ref{VGtimesFminusVApprox}), it's possible to take $c_{s0} =1$ and still obtain a well defined result. Doing so results in the following for the functions $g$, $f$ and $V$:
\begin{gather}
g(x) 
\approx
g_{0}
+
\frac{\epsilon \H_{0}^{2}}{f_{0}}
\frac{x-x_{0}}{x_{0}}
,
\quad
f(\varphi)
\approx
\frac{f_{0}}{(1 + \frac{\epsilon \H_{0}}{\dot{\varphi}_{0}} (\varphi - \varphi_{0}))^{2}}
,
\quad
V(\varphi)
\approx
\frac{f_{0} g_{0} - (2 \epsilon - 3) \H_{0}^{2}}{(1 + \frac{\epsilon \H_{0}}{\dot{\varphi}_{0}} (\varphi - \varphi_{0}))^{2}},
\label{gfVinCanonicalLim}
\end{gather}
where now $\H_{0}^{2} = \epsilon$ since $c_{s0} = 1$. We refer the reader to fig. \ref{FVersusPhiForFtimesGMinusV} for a comparison of $f(\varphi)$ in (\ref{gfVinCanonicalLim}) and $f(\varphi)$ for general values of $\omega$ and $c_{s0}$. The action in the $c_{s0} \rightarrow 1$ limit when $\omega = 0$ is
\begin{gather}
\varrho(\varphi,x)
=
\frac{\epsilon \H_{0}^{2}}{(1 + \frac{\epsilon \H_{0}}{\dot{\varphi}_{0}} (\varphi - \varphi_{0}))^{2}}
\frac{x}{x_{0}}
-
\frac{(3 - \epsilon) \H_{0}^{2}}{(1 + \frac{\epsilon \H_{0}}{\dot{\varphi}_{0}} (\varphi - \varphi_{0}))^{2}}.
\label{UniqueAction}
\end{gather}
This action can be related to the canonical action (\ref{ExpCanonicalAction}) through the field redefinition defined by
\begin{gather}
1 + \frac{\epsilon \H_{0}}{\dot{\varphi}_{0}} (\varphi - \varphi_{0})
=
e^{\sqrt{\frac{\epsilon}{2}} (\tilde{\varphi} - \tilde{\varphi}_{0})}.
\end{gather}
Under this redefinition, the new action is
\begin{gather}
\varrho(\tilde{\varphi},\tilde{x})
=
\tilde{x}
-
(3 - \epsilon)
\H_{0}^{2}
e^{- \sqrt{2 \epsilon} (\tilde{\varphi} - \tilde{\varphi}_{0})}.
\end{gather}
This is the same as the canonical action (\ref{ExpCanonicalAction}) found in the first example. To ensure a smooth transition to the canonical action we must separate the kinetic function as we did in (\ref{gSep}) so that $G(0) = 0$ in the $\omega \rightarrow 0 $ and $c_{s0} \rightarrow 1$ limits. Upon inspection of $g(x)$ in (\ref{gfVinCanonicalLim}) we see that the redefined kinetic function $G(x)$ is
\begin{gather}
G(x)
=
g(x)
-
\frac{f_{0} g_{0} - \epsilon \H_{0}^{2}}{f_{0}}.
\end{gather}
In doing so the potential is redefined according to (\ref{RedefPotential}) as  
\begin{gather}
v(\varphi)
=
V(\varphi)
-
\frac{g_{0} f_{0} - \epsilon \H_{0}^{2}}{f_{0}}
f(\varphi).
\label{VShift}
\end{gather}
There is a subtlety in this analysis that should be addressed. In order to reclaim the canonical action we needed to take the limits $\omega \rightarrow 0$ and $c_{s0} \rightarrow 1$ simultaneously. In our case we took that limit by setting $\omega = 0$ and then letting $c_{s0}$ approach one. However, this is by no means the only way to take the limit. For example, we could have approached the limit by setting $\omega = 1 - c_{s0}^{2}$ and then take the limit as $c_{s0}$ goes to one. Had we taken the limit from a different direction it is possible that  the action that resulted could have been different from the canonical action (\ref{ExpCanonicalAction}). After some inspection, it can be shown that under a field redefinition $\varphi = h(\tilde{\varphi})$ such that $f^{-1}(\varphi) = (h^{\prime}(\tilde{\varphi}))^{2}$, the potential in the canonical limit is given by
\begin{gather}
V(\tilde{\varphi})
=
(3 -  \epsilon )
\H_{0}^{2}
e^{- \frac{2 \epsilon \H_{0}}{\dot{\varphi}_{0}} (\tilde{\varphi} - \tilde{\varphi}_{0}) }
+
\left(
g_{0} f_{0}
-
\epsilon \H_{0}^{2}
\right)
\frac{f(\tilde{\varphi})}{f_{0}}.
\end{gather}
Here $f(\tilde{\varphi}) = f(h(\tilde{\varphi}))$ is the function $f$ when the canonical limit is taken. It is simple to show that the canonical limit of $f$ is not unique, which means that the potential is also not unique. However, if we redefine our potential according to (\ref{VShift}) instead, the new potential $v$ is unique, and the action that results is canonically equivalent to (\ref{ExpCanonicalAction}).

\subsubsection{Case 3: $f_{1}(\varphi)$ and $f_{2}(\varphi)$ Unknown}
\label{f1f2Unknown}
We now bring up a case that will be of particular interest to reconstructions of the DBI action. We start by assuming that the action $\varrho(\varphi,x)$ has the form
\begin{gather}
\varrho(\varphi,x)
=
P(f_{1}(\varphi),
f_{2}(\varphi), x),
\end{gather}
where $f_{1}$ and $f_{2}$ are unknown functions of $\varphi$. Unlike the previous cases, the functional dependence of the action with respect to $x$ is assumed to be known exactly. In this case it is possible to obtain a set of algebraic equations of the two unknowns $f_{1}$ and $f_{2}$. The first of these equations can be most easily obtained by going back to the original definition of the sound speed (\ref{csInTermsOFgamma}):
\begin{gather}
c_{s}^{2}
=
\frac{p_{,x}}{2 x p_{,xx} + p_{,x}}
\quad
\Rightarrow
\quad
P_{,xx}
=
\frac{1}{2 x}
\left(
\frac{1}{c_{s}^{2}}
-1
\right)
P_{,x}.
\label{SoundSpeedf1f2Unknown}
\end{gather}
This equation together with (\ref{Xequation}) and the Friedmann equation (\ref{FriedmannThree}) are enough to find $f_{1}(\varphi)$ and $f_{2}(\varphi)$ in terms of the observables. In the next section we will see explicitly how the equations (\ref{SoundSpeedf1f2Unknown}), (\ref{Xequation}) and (\ref{FriedmannThree}) come together to reconstruct the DBI action from the power spectrum data.

\section{DBI inflation}
\label{DBIInflation}
In realistic string and M-theories, the number of space-time dimensions is 10 or 11 dimensions. The extra $6$ or $7$ dimensions are compactified to small sizes, leaving the effective theory at low energies a theory of physics in four dimensions. The various moduli that control the shape (complex structure moduli) and size (K\"ahler moduli) of the internal space, also determine the nature of the four-dimensional low-energy effective theory. Therefore, fixing these moduli is an important step in establishing a connection between string theory and the standard model. In recent years, much attention has been paid to flux compactifications as a potential means of stabilizing string moduli\footnote{For a review of flux compactifications see \cite{Grana:2005jc}.}. In a flux compactification, various fluxes wrap around closed cycles in the internal manifold creating a potential for the complex structure moduli. The best known of the these takes place in type IIB string theory. Here the internal space is six-dimensional Calabi-Yau and the 3-form fluxes $F_{3}$ and $H_{3}$ create a superpotential that fixes the complex structure \cite{Giddings:2001yu}. These 3-form fluxes source a warping of the geometry of the internal manifold. In the type IIB flux compactification, the ansatz of the line element is taken as
\begin{gather}
ds_{10}^{2}
=
h^{-1/2}(y) g_{\mu\nu} dx^{\mu} dx^{\nu}
+
h^{1/2}(y) g_{mn} d y^{m} dy^{n}.
\label{10DMetric}
\end{gather}
Here $h$ is the warp factor which is sourced by the fluxes and varies only along the dimensions of the internal manifold. In DBI inflation, which we will be considering in this section, the local geometry of the internal manifold is a Klebanov-Strassler throat geometry \cite{Klebanov:2000hb}, and is described by the metric
\begin{gather}
g_{mn} d y^{m} dy^{n}
=
dr^{2} + r^{2} ds^{2}_{X_{5}}.
\end{gather}
Here $ds^{2}_{X_{5}}$ is the line element of a five-dimensional manifold $X_{5}$, which forms the base of the KS throat. The coordinate $r$ runs along depth of the throat. For our purposes we will only consider motion along $r$ and integrate over the base manifold $X_{5}$\footnote{Fluctuations of the brane position along the transverse directions of the KS throat have been mentioned as a possible source of entropy perturbations \cite{Easson:2007dh,Brandenberger:2007ca}. These could serve as a further constraint on the form of the action.}. The warping of the internal space creates a natural realization of the Randall-Sundrum model \cite{Randall:1999ee}, and has also provided model builders with a new approach to developing string theory based models of inflation. The most popular inflation model that makes use of this warping is DBI inflation, which is the primary focus of this section.

In the simplest DBI inflation models \cite{Silverstein:2003hf,Alishahiha:2004eh,Chen:2005ad} a $D3$ brane travels along the $r$ direction, either into or out of the KS throat. The $D3$ brane extends into the $3$ non-compact space dimensions and is point like in the internal manifold. The standard DBI action for the $D3$ brane is  
\begin{gather}
S_{DBI}
=
-
\int d^{4}x
\sqrt{-g}
\left[
f^{-1}(\phi) \sqrt{1 - 2 f(\phi) X}
-
f^{-1}(\phi)
+
V(\phi)
\right].
\label{DBI Action}
\end{gather}
Here $\phi = \sqrt{T_{3}} r$ (where $T_{3}$ is the $D3$ brane tension) is a rescaling of the coordinate $r$ and will play the role of the inflaton. The quantity $f^{-1} = T_{3} h^{-1}$ is the rescaled warp factor. The metric $g$ that appears in (\ref{DBI Action}) is the metric on the $3+1$ dimensional non-compact subspace which describes the geometry of our familiar 4 dimensional space-time. We will continue to assume that the geometry of the $3+1$ dimensional subspace is described by the FRW metric with zero curvature. The energy density and pressure in the non-compact subspace due to the brane are given by 
\begin{gather}
\rho
=
f^{-1}
\left(\gamma 
-
1\right)
+
V,
\\
p 
=
(\gamma f)^{-1}
\left(\gamma
-1
\right)
-
V.
\end{gather}
Here $\gamma$ is a new parameter, not found in the standard canonical inflation. In terms of the quantities in the DBI action, $\gamma$ is defined as
\begin{gather}
\gamma = \frac{1}{\sqrt{1 - 2 f(\phi) X}}.
\label{DBIDefinitionofGamma}
\end{gather}
The $\gamma$ defined here is analogous to the Lorentz factor in special relativity, and will hence-forth be referred to as the Lorentz factor. The Lorentz factor places an upper limit on the speed of the brane as it travels through the KS throat. Since the kinetic energy of the brane is limited, this allows one to get a sufficient amount of inflation even with potentials that would be considered too steep to use in standard canonical inflation. 

In our study of the DBI model we will be assuming that the scalar and tensor spectra are approximately (\ref{PscalarApproxOne}) and (\ref{PtensorApproxOne}), respectively. With these as our inflationary observables, we found that $\epsilon$ was constant (\ref{Epsilon}). The fact that $\epsilon$ is a constant indicates that inflation will not end on it's own, and instead some other mechanism such as $D3$-$\overline{D3}$ annihilation \cite{Barnaby:2004gg} must be used to provide a graceful exit. Since our study is concerned more with the physics during inflation, this topic will not be addressed further. We will now present a generalized DBI action, and show how it is reconstructed from the inflationary observables.

\subsection{A Generalized DBI Model}
Having sketched out the general method for reconstructing different types of inflationary actions in section \ref{EqusofInflation}, it is now time to apply these methods to a DBI-type action given by
\begin{gather}
\varrho(\varphi,x)
=
P(x,\F(\varphi),\V(\varphi)),
\label{GeneralizedDBIAction}
\end{gather}
where
\begin{gather}
P(z_{1},z_{2},z_{3})
=
-z_{2}^{-1}
\left(
\sqrt{1 - 2 z_{2} z_{1}}
-1
\right)
-
z_{3}
\label{GenDBIActionPFunc}.
\end{gather}
Here $\F(\varphi) = \alpha^{2} M_{pl}^{2} f(\varphi)$ is the (dimensionless) warp factor in the throat, and $\V(\varphi) = (\alpha M_{pl})^{-2} V(\varphi)$ is the (dimensionless) potential. In the KS throat geometry the warp factor is taken to be $\F \propto \varphi^{-4}$. The potential $\V$ is assumed by many to be quadratic in $\varphi$. For the purposes of this study we will not assume a priori any form for the functions $\F$ and $\V$, and instead allow the inflationary observables to determine them. Now that we have established the general form of the action, we can use the procedure outlined in section \ref{f1f2Unknown} to find $\F$ and $\V$. Turning to equations (\ref{SoundSpeedf1f2Unknown}) and (\ref{Xequation}) we find the relations
\begin{gather}
\frac{P_{,xx}}{P_{,x}} = \frac{\F}{1 - 2 \F x}
=
\frac{1}{2 x} 
\left(
\frac{1}{c_{s}^{2}}
-1
\right),
\qquad
x = \epsilon \H^{2} \sqrt{1 - 2 \F x}.
\end{gather}
Solving for $\F$ and $x$: 
\begin{gather}
\F(\varphi) = \frac{1 - c_{s}^{2}}{2 \epsilon \H^{2} c_{s}}
,
\qquad
x
=
\epsilon \H^{2} c_{s}.
\label{FEqu}
\end{gather}
Comparing the second equation above with (\ref{Xequation}) and recalling the definition of (\ref{DBIDefinitionofGamma}), we find that 
\begin{gather}
c_{s} = \frac{1}{\gamma}. 
\label{CsRelatedtoGammaDBI}
\end{gather}
This result is characteristic of DBI inflation and holds regardless of the warp factor and potential used. Having found $\F$, equation (\ref{FriedmannThree}) tells us what $\V$ is:
\begin{gather}
\V(\varphi)
=
3 \H^{2}
+
\frac{1}{\F}\left(\frac{1}{c_{s}} - 1\right).
\end{gather}
Having already found the expression for $\F$ in (\ref{FEqu}), we can now write down the important  reconstruction equations for $\V$ and $\F$ in terms of $\H$, $c_{s}$ and  $\epsilon$:
\begin{gather}
\V(k)
=
\H^{2}
\left(
3
-
\frac{2 \epsilon}{1 + c_{s}}
\right),
\label{VinTermsofPhandR}
\\
\F(k)
=
\frac{1
-
c_{s}^{2}}{2 \epsilon c_{s} \H^{2}}.
\label{FinTermofPhandR}
\end{gather}
To turn $\F$ and $V$ into functions of $\varphi$ we need to integrate our solution for $x$ (\ref{FEqu}) to find $\varphi$. Taking equation (\ref{Xequation}) to find an expression for $\dot{\varphi}$, we find that in the case of DBI inflation
\begin{gather}
\dot{\varphi}
=
\pm
\sqrt{2 \epsilon c_{s}} \H.
\label{DBIphiEquation}
\end{gather}
The sign of the right hand side of the equation is ambiguous, due to the square root taken to get this equation from (\ref{Xequation}). The sign is left arbitrary for now and will be specified later based on the requirement that $\varphi$ be positive. Once we solve for $\varphi(\tau)$ in (\ref{DBIphiEquation}) and invert to get $\tau(\varphi)$, we can then find a solution for $\V(\varphi)$ and $\F(\varphi)$. Now that we have laid the ground work for generating the functions of the generalized DBI action, the next section will show how the perturbation spectra are used to obtain explicit expressions for $\F(\varphi)$ and $\V(\varphi)$.

\subsection{The Warp Factor and Potential in DBI Inflation}
\label{WithoutRunning}
In this section we will now use the program that was laid out at the end of the previous section to find an exact solution for the warp factor and potential in the action (\ref{GeneralizedDBIAction}). We will again assume that the scalar and tensor spectra have a power-law dependence with respect to the scale $k$. Therefore, the sound speed, Hubble parameter and $\epsilon$ are the same as those found in section \ref{Case1}. Thus, the potential as a function of time is
\begin{gather}
\V
=
\frac{\H_{0}^{2}}{(1 + \epsilon \H_{0} (\tau -\tau_{0}))^{2}}
\left(
3
-
\frac{2\epsilon}{1 + \frac{\H_{0}^{2}}{\epsilon} (1 + \epsilon \H_{0} (\tau - \tau_{0}))^{-\omega} }
\right),
\label{PotentialTSol}
\end{gather}
and the warp factor as a function of time is
\begin{gather}
\F
=
\frac{(1 + \epsilon \H_{0} (\tau - \tau_{0}))^{\omega + 2}}{2 \H_{0}^{4}}
\left(
1
-
\frac{\H_{0}^{4}}{\epsilon^{2}}
\left(
1 + \epsilon \H_{0} (\tau - \tau_{0})
\right)^{- 2 \omega}
\right).
\label{WarpFactorTSol}
\end{gather}
Since we want inflation to occur we need $\epsilon$ to be such that $\epsilon <1$, which according to equation (\ref{Epsilon}) implies that $n_{T} \leq 1$. However, $n_{T} = 1$ can be dismissed as physically unreasonable since that would imply $c_{s0} \propto \frac{1}{\epsilon} \rightarrow \infty$. After substituting our solutions for $\H(\tau)$ and $c_{s}(\tau)$ into the equation of motion for $\varphi(\tau)$ we get
\begin{gather}
\dot{\varphi}
=
\pm \sqrt{2} \H_{0}^{2} (1 + \epsilon \H_{0} (\tau - \tau_{0}))^{-\frac{\omega}{2} - 1}.
\label{PhiDK}
\end{gather}
Once we integrate this expression we can obtain an answer for $\varphi(\tau)$. As a matter of convenience we will set the value of the integration constant that results to zero. Later, once we have found $\F$ and $\V$, we will see that this choice allows for a correspondence between the reconstructed functions and their theoretically derived counterparts. After integrating (\ref{PhiDK}) we find that
\begin{gather}
\varphi(\tau) 
=
\int^{\tau}
\dot{\varphi}
d\tau
=
\mp \frac{2 \sqrt{2} \H_{0}}{\epsilon \omega}
(1 + \epsilon \H_{0} (\tau - \tau_{0}) )^{- \omega / 2}.
\label{PhiKntLESSns}
\end{gather}
Since we are interested in eventually connecting the reconstructed action with the standard DBI model we need to keep the inflaton, which is just a rescaled radial coordinate, positive. The sign that we choose in (\ref{PhiKntLESSns}) will therefore depend on the sign of $\omega$. We can write the general solution as
\begin{gather}
\varphi(\tau)
=
\varphi_{0}
(1 + \epsilon \H_{0} (\tau - \tau_{0}) )^{- \omega / 2}
\end{gather}
where 
\begin{gather}
\varphi_{0}
=
2 \sqrt{2}
\left|\frac{\H_{0}}{\epsilon \omega}\right|
\end{gather}
In the case where $\omega >0$ the field $\varphi$ decreases monotonically to zero as time passes, which implies that the brane is falling into the throat. This corresponds to the UV DBI scenario. If on the other $\omega < 0$, then $\varphi$ increases monotonically with time and the brane falls out of the throat, which corresponds to IR DBI inflation. Solving for time in terms of $\varphi$
\begin{gather}
1 + \epsilon \H_{0} (\tau - \tau_{0})
= 
\left(
\frac{\varphi}{\varphi_{0}}
\right)^{- \frac{2}{\omega}}.
\end{gather}
Substituting this for $1 + \epsilon \H_{0} (\tau - \tau_{0})$ in the expressions we found for the potential and warp factor we find that $\V$ as a function of $\varphi$ is  
\begin{gather}
\V
=
\H_{0}^{2}
\left(
\frac{\varphi}{\varphi_{0}}
\right)^{\frac{4}{\omega} }
\left(
3
-
\frac{2 \epsilon}{1 + \frac{\H_{0}^{2}}{\epsilon}  \left(
\frac{\varphi}{\varphi_{0}}
\right)^{2}}
\right),
\end{gather}
and the warp factor as a function of $\varphi$ is
\begin{gather}
\F
=
\frac{1}{2 \H_{0}^{4}}
\left(
\frac{\varphi}{\varphi_{0}}
\right)^{-2 - \frac{4}{\omega}}
\left[
1
-
\frac{\H_{0}^{4}}{\epsilon^{2}}
\left(
\frac{\varphi}{\varphi_{0}}
\right)^{4}
\right].
\label{FinTermsOfPhi}
\end{gather}
Furthermore, when $\gamma$ is expressed as a function of $\varphi$, it takes on a very simple form:
\begin{gather}
\gamma 
= 
\gamma_{0} \left(\frac{\varphi_{0}}{\varphi}\right)^{2}
=
\frac{8}{\epsilon \omega^{2}}
\frac{1}{\varphi^{2}},
\label{DBIGamma}
\end{gather}
where $\gamma_{0} = \frac{1}{c _{s0}}$. It is interesting to note that (\ref{DBIGamma}) is the same as the approximate results found in the theoretically inspired DBI model \cite{Alishahiha:2004eh}. The potential and warp factor derived here are the same as those found in \cite{Chimento:2007es}. There the authors reconstructed the potential and warp factor by assuming that the equation of state $w = \frac{p}{\rho}$ was a constant and that $\varphi \propto \tau^{- \omega /2}$. In contrast, we have reconstructed the potential and warp factor under the assumption that the scalar and tensor perturbations are (\ref{PscalarApproxOne}) and (\ref{PtensorApproxOne}). The non-gaussianity in this DBI model is the same result that one comes across in the literature \cite{Chen:2006nt}:
\begin{gather}
f_{NL}
=
\frac{35}{108}
\left(
\frac{1}{c_{s}^{2}}
-
1
\right).
\label{DBIfNL}
\end{gather}
This particularly simple result is a general feature of DBI, and is independent of the warp factor and potential. This result for the non-gaussianity (\ref{DBIfNL}) also follows from consistency of the reconstruction equations (\ref{SoundSpeedEqu}) and (\ref{NonGaussEqu}). Thus, (\ref{DBIfNL}) can be viewed as a consistency relation analogous to those found in (\ref{fTimesgConsistency}) and (\ref{gVRelation}). An interesting generalization to consider is 
\begin{gather}
\varrho(\varphi,x)
=
P(g(x),\F(\varphi),\V(\varphi))
\end{gather}
where $P(z_{1},z_{2},z_{3})$ is defined by (\ref{GenDBIActionPFunc}). As a consistency check, one can easily show that if the non-gaussianity is equal to (\ref{DBIfNL}), and the potential and warp factor are given by (\ref{PotentialTSol}) and (\ref{WarpFactorTSol}), respectively, then $g(x) = x$ is a solution to the reconstruction equations (\ref{SoundSpeedEqu}) and (\ref{NonGaussEqu}).

Having found the potential and warp factor as functions of the inflaton, we can now say that our task is at an end. Amazingly enough, despite the complicated form of the reconstruction equations an exact solution for $\V$ and $\F$ was available even for semi-realistic scalar and tensor spectra. In the next section we will discuss the properties of the reconstructed action, and its correspondence with the theoretically derived action of DBI inflation.

\subsection{Discussion}
In section \ref{Case1} we found that not all values of the spectral indices lead to inflationary and/or physically sensible actions. Specifically, we showed that unless $n_{T} < 1$ the matter described by the action was unable to drive an inflationary phase. Furthermore, when this constraint on $n_{T}$ was considered in conjunction with the requirement that the sound horizon decrease as inflation occurs, the scalar spectral index had to be bounded like $n_{s} < 2$. These results hold for any reconstructed action that was derived assuming the observational inputs (\ref{PscalarApproxOne}) and (\ref{PtensorApproxOne}). However, even if these constraints are satisfied, it is not guaranteed that the reconstructed action is physically sensible when interpreted in the context of a given theoretical construction. For example, if we are to interpret the action reconstructed in this section as a DBI action, then $\gamma > 0$. Doing so would be contrary to its definition (\ref{DBIDefinitionofGamma}) within the context of DBI inflation. In this case, since $\gamma = \frac{1}{c_{s}}$ the fact that we already have enforced the constraint $c_{s} >0$ automatically keeps $\gamma$ positive. We will see later, however, that the constraints found in \ref{Case1} are not sufficient for our reconstructed action to be interpreted as a DBI action.  

First, let's consider the warp factor (\ref{FinTermsOfPhi}) and what constraints it places on the observables. Suppose $n_{T} < n_{s}$. In this case the exponent of $\varphi$ in front of the square brackets in (\ref{FinTermsOfPhi}) will always be negative. Therefore, for small values of $\varphi$ the leading order behavior of $\F$ will go like $\varphi^{-a}$ where $a > 0$. Thus, the warp factor increases as we fall into the throat, which is what we would expect for a warped compactification in string theory. On the other hand, if $n_{T} > n_{s}$ the leading order behavior of $\F$ will be $\varphi$ to some positive or negative power, depending on the relative difference between $n_{s}$ and $n_{T}$. If the difference between $n_{T}$ and $n_{s}$ is too small, then to leading order, $\F$ will scale like $\varphi$ to some positive power. This indicates that the warp factor gets smaller as we reach the bottom of the throat, which is a scenario that is difficult to embed into a string theory compactification. However, If the difference between $n_{s}$ and $n_{T}$ is large enough, then it is possible to get a more sensible solution where $\F \sim \varphi^{-a}$. In general, the condition that $\F$ increases as we approach the bottom of the throat implies that 
\begin{gather}
-1
-
\frac{2}{\omega}
< 0
\quad
\Rightarrow
\quad
\omega < -2 
\,\,\,
\textrm{ or }
\,\,\,
\omega>0
.
\label{DivFCondition}
\end{gather}
The regions in the $n_{s}$-$n_{T}$ parameter space where the condition (\ref{DivFCondition}) is satisfied are shown in
\begin{figure}
\begin{center}
\includegraphics[scale=.5]{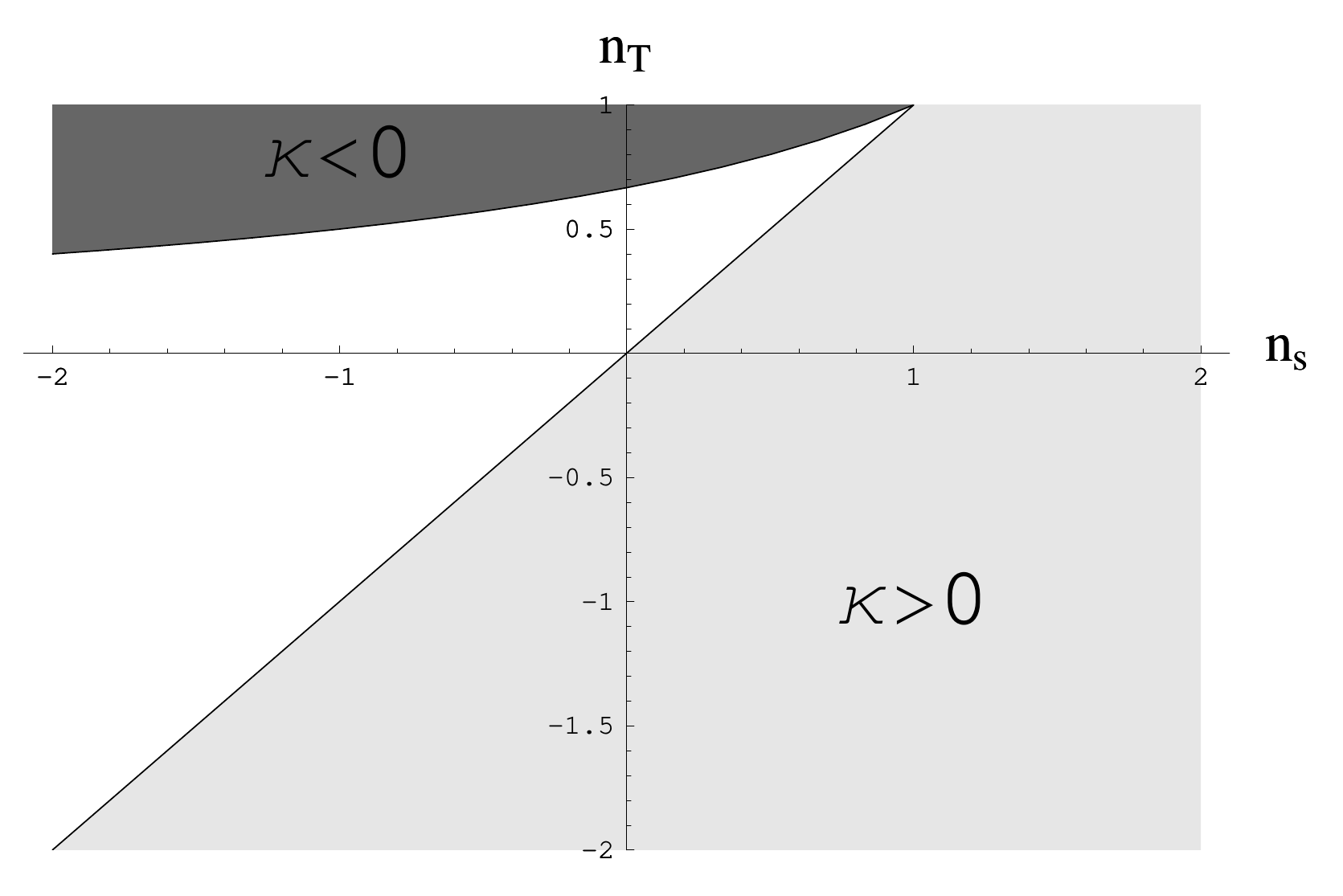}
\caption{Plot depicting the regions of the $n_{s}$-$n_{T}$ parameter space such that $\F$ diverges at the origin. The region in the upper left hand corner is the region of IR DBI inflation while the corner at the bottom right hand corner corresponds to UV DBI. Only those values of $n_{s}$ and $n_{T}$ with $n_{s} <2$ and $n_{T} <1$ were considered, since any point outside that region would lead to an unphysical and/or non-inflationary action.}
\label{UVandIRRegions}
\end{center}
\end{figure}
fig. \ref{UVandIRRegions}. The region shown in light grey in fig. \ref{UVandIRRegions} is defined by $\omega>0$, or equivalently $n_{T} < n_{s}$. This region corresponds to the UV phase of DBI inflation. The region in dark grey is defined by $\omega<-2$, or equivalently $n_{T} > - \frac{2}{n_{s}-3}$, and corresponds to the IR phase. As fig. \ref{UVandIRRegions} illustrates, if $n_{s}$ is restricted to the presently favored value $n_{s}\sim 0.96$, then the value of $n_{T}$ is tightly constraint in the IR region but relatively unrestricted in the UV phase. Furthermore, since $n_{T} <1$, only those models in the UV phase can have blue-tilt. Recent CMBR data favors a blue-tilted spectrum, but only if there is a running spectral index \cite{Komatsu:2008hk}. Although running spectral indices would be an interesting extension of this analysis, we will leave this topic to future studies.

An unpleasant feature of the warp factor (\ref{FinTermsOfPhi}) is that it becomes negative when $\varphi > \varphi_{0} \H_{0} / \sqrt{\epsilon}$. This is particularly distasteful since the metric (\ref{10DMetric}) depends on $f^{1/2}$, which means that at sufficiently large $\varphi$ the metric is imaginary. The values of $\tau$ where the warp factor is positive are given by
\begin{gather}
\tau - \tau_{0}
>
\frac{c_{s0}^{\frac{1}{\omega}} - 1}{\epsilon\H_{0}}
\qquad
\textrm{for }\omega > 0
\,\,\, \textrm{(UV)},
\nn \\
\tau - \tau_{0}
<
\frac{c_{s0}^{\frac{1}{\omega}} - 1}{\epsilon\H_{0}}
\qquad
\textrm{for }\omega < 0
\,\,\, \textrm{(IR)}.
\label{tbounds}
\end{gather}
One can show using equation (\ref{logkInTermsofTime}) that this is equivalent to the bounds in (\ref{CsKbounds}). This is no coincidence; it is a result of the fact that $\F$ is proportional to $1 - c_{s}^{2}$ (\ref{FEqu}). Therefore, for the same reasons that were explained in section \ref{Case1}, the perturbations (\ref{PscalarApproxOne}) and (\ref{PtensorApproxOne}) can only be  used as approximations. The inequalities in (\ref{tbounds}) tell us that $\F$ is a valid warp factor towards the end of inflation in the case of UV DBI, and at the beginning of inflation in IR DBI. The time $\tau_{0}$ at which the initial conditions are specified should be at the beginning of inflation in the UV scenario, and at the end in the case of IR DBI. If we choose $\tau_{0}$ in this manner then the approximations for the perturbation spectra (\ref{PscalarApproxOne}) and (\ref{PtensorApproxOne}) will lead to realistic sound speeds for the entire duration of the inflationary episode.

It is worth asking if the $\F$ that we have derived in (\ref{FinTermsOfPhi}) can approximate the AdS warp factor derived from theory. It is clear from (\ref{FinTermsOfPhi}) that this can be achieved if and only if $\omega = 2$. However, the only way we can get $\omega =2$ is if either i.) $n_{T} = 2$ ii.) $n_{T} \rightarrow \infty$ or iii.) $n_{s} =1$. As we have already seen, case i.) is unphysical, and case ii.) is difficult to imagine taking place. While case iii.) is unlikely to be true exactly, it is nevertheless the more realistic of the three, especially when you consider that observation suggests that $n_{s} \approx 0.96$. If we do set $n_{s} =1$, the warp factor and potential become
\begin{gather}
\F(\varphi)
=
\frac{2}{\epsilon^{4}}
\frac{1}{\varphi^{4}}
-
\frac{1}{2 \epsilon^{2}},
\end{gather}
\begin{gather}
\V
=
\frac{3\epsilon^{2}}{2}
\varphi^{2}
-
\frac{\epsilon^{3} \varphi^{2}}{1 + \frac{\epsilon}{2} \varphi^{2}}.
\end{gather}
In the case where the field range is small\footnote{In \cite{Baumann:2006cd} it was shown that DBI inflation is only consistent when the magnitude of the inflaton field is sub-planckian: $\varphi \ll 1$.} these are approximately
\begin{gather}
f(\phi)
\approx
\frac{2M_{pl}^{2}}{\alpha^{2} \epsilon^{4}}
\frac{1}{\phi^{4}},
\label{ApproxWarpFactor}
\end{gather}
\begin{gather}
V(\phi)
\approx
\frac{\epsilon^{2} \alpha^{2} (3 - 2 \epsilon) }{2}
\phi^{2}.
\end{gather}
Where we have reverted back to the standard, dimensionful $f$, $V$ and $\phi$ for clarity's sake. It is a bit of a surprise that it in process of trying to recover the AdS warp factor we have stumbled upon the commonly used potential in UV DBI inflation. If we take (\ref{ApproxWarpFactor}) and demand that it is consistent with the theoretical result we can arrive at a condition on $\epsilon$ in terms of the D3 charge. Recall that in the KS throat $f(\phi)$ is given by 
\begin{gather}
f(\phi)
=
\frac{2 T_{3} R^{4}}{\phi^{4}},
\end{gather}
where $T_{3} =\frac{1}{(2 \pi)^{3}} \frac{1}{g_{s} (\alpha^{\prime})^{2}}$ and 
\begin{gather}
R^{4}
=
4 \pi g_{s} N (\alpha^{\prime})^{2} \frac{\pi^{3}}{\textrm{Vol}(X_{5})}.
\end{gather}
Consistency with (\ref{ApproxWarpFactor}) demands that
\begin{gather}
\frac{2 M_{pl}^{2}}{\alpha^{2} \epsilon^{4}}
=
\frac{\pi N}{\textrm{Vol}(X_{5})}
\quad
\Rightarrow
\quad
\epsilon \approx
\frac{10^2}{N^{1/4}}.
\end{gather}
In order to get an inflationary phase $N\approx 10^{10}$, putting us well within the range of validity for the supergravity approximation. While it is interesting that the standard D3 brane DBI model can be recovered from a near scale invariant scalar power spectrum, it has been acknowledged that this inflation model is problematic. In \cite{Baumann:2006cd} Baumann and McAllister found that while present bounds on non-gaussianity imply that $N \lesssim 38$, primordial perturbations imply that $N \gtrsim 10^{8} \textrm{Vol}(X_{5})$. These two limits are incompatible unless $\textrm{Vol}(X_{5}) \lesssim 10^{-7}$. It is not clear that such a space could be naturally embedded into a string theory compactification. More general warp factors and potentials have been considered in \cite{Peiris:2007gz}. There it was found that models could not simultaneously satisfy bounds on the field range and observational bounds on the non-gaussianity. Therefore, even though our warp factor and potential matches the theoretically based predictions, the problems inherent in the DBI model carry over into its generalizations.

\section{Conclusion}
\label{Conclusion}
In this paper we have presented a method for deriving the actions of single field inflation models using CMBR data. This method allows one to derive up to three unknown functions of the action using the scalar perturbation $P_{s}$, tensor perturbation $P_{t}$ and the non-gaussianity $f_{NL}$. After stating the reconstruction equations, we carried out the reconstruction procedure for two simple examples. For the purposes of the reconstruction, we assumed that the scalar and tensor spectra were power-law dependent on the scale $k$, with the spectral indices kept as free parameters. In the first example we assumed that the action had the form shown in equation (\ref{gVAction}), and used the reconstruction equations to obtain the action as a function of the spectral indices. In this example there were only two unknown functions, thus the reconstruction equations also lead to a consistency relation (\ref{gVRelation}) between the $f_{NL}$, $c_{s}$ and the slow roll parameters. However, this consistency relation is only well defined when the sound speed is not a constant. 

In the second example, the action depended on three unknown functions and therefore required all three reconstruction equations. In order to simplify the discussion we took as our input for the non-gaussianity $f_{NL} =0$. Although we were unable to express the action in terms of elementary functions we were able to obtain the action numerically and approximately assuming $c_{s0} \approx 1$ and $\omega \approx 0$. We showed that the action in this example was canonically equivalent to the canonical action derived at end of the previous section. In discussing this example we also pointed out possible ambiguities in the program relating to how one defines a separation between the kinetic and potential terms. 

In section \ref{DBIInflation}, we used the procedure to derive and study the warp factor and potential in a generalized DBI inflation model. Again, we assumed that both of the perturbation spectra scaled like $k$ to some power. Exact expressions for the warp factor and potential were then derived, each having an explicit dependence on the spectral indices. The demand for a physically sensible DBI inflation model placed constraints on the spectral indices. In addition we found that the derived action approximates the original UV DBI inflation model in the case where $\phi \ll M_{pl}$. Unfortunately, the problems that have plagued UV DBI inflation are still present in our case.

This procedure was shown to be useful in studying how the action of a general inflation model depends on the observables. For example, we found that if the scalar and tensor perturbation spectra went like $k$ to an arbitrary power, the reconstruction would lead to a realistic inflationary model only if $n_{T} <1$ and $n_{s} <2$.  Furthermore, to keep the speed of fluctuations from becoming superluminal, the range of $k$ over which the approximations for the spectra (\ref{PscalarApproxOne}) and (\ref{PtensorApproxOne}) are taken, had to limited. When we reconstructed a generalized DBI action in section \ref{DBIInflation}, further constraints were needed to keep the action compatible with an interpretation of DBI inflation. Specifically, we found that in order to keep the warp factor positive, the field range had to be limited. Furthermore, in the theoretically motivated DBI model, the warp factor increases as we reach the bottom of the warped throat. In order for this to be true in our reconstruction, the spectral indices needed to satisfy the additional constraints: $\omega < -2$ or $\omega >0$.

In this paper we have only considered the simplest of the DBI inflation models, which unfortunately suffers from several inconsistencies. However, there are many extensions of the D3 brane DBI model that can circumvent some of the problems of the original. Some of these extensions include using wrapped D5 branes \cite{Becker:2007ui,Kobayashi:2007hm}, multiple D3 branes \cite{Thomas:2007sj}, and multiple throats \cite{Chen:2004gc}. Each of these models has its potential advantages and drawbacks. Applying our reconstruction procedure may help to further elucidate their relative strengths and weaknesses. Furthermore, we have limited ourselves to perturbations with simple power law behavior. However, this naive assumption may be incorrect. It is easy to imagine that the spectral indices themselves are also scale dependent. Based on the results of this paper we can predict what kind of effect a running spectral index would have on the physics of the underlying models. For instance, in the generalized DBI model it is possible for the spectral indices to change during inflation in such a way as t o pass from the IR to the UV phase\footnote{Fig. \ref{UVandIRRegions} implies that inflation can change between UV and IR phases only if it passes through the exactly scale-invariant point: $(n_{s},n_{T}) = (1,1)$.}. Transition between phases would correspond to a completely different physical scenario, one where the brane falls out of one throat and back into another. Therefore, running spectral indices would describe multi-throat DBI inflation. A model which has so far been shown to be internally consistent \cite{Bean:2007eh}.  

This study has also raised some other questions that may be worth investigation. In particular what is the relation between actions that yield the same observables. It may be possible to define a group of transformations which leave the perturbation spectra and the non-gaussiantiy invariant. Such a set of transformations would allow us to classify actions based on the observables they yield. Another interesting possibility that came out of this study is the idea of using the reconstruction equations as a way to generate consistency relations between $f_{NL}$, the sound speed $c_{s}$ and the slow roll parameters. These questions will be left for future studies.

\section{Acknowledgements} We would like to thank Louis Leblond and Dragan Huterer for helpful discussions. We would also like to give a special thanks to Daniel Chung for his insightful comments on our paper.

\bibliographystyle{utcapsMyFix}
\bibliography{Inflation-Inverse-Paper-Bib}

\providecommand{\href}[2]{#2}\begingroup\raggedright\begin{thebibliography}{10}

\bibitem{Smoot:1992td}
G.~F. Smoot {\em et al.}, ``{Structure in the COBE differential microwave
  radiometer first year maps},'' {\em Astrophys. J.} {\bf 396} (1992)
L1--L5.

\bibitem{Grishchuk:1974ny}
L.~P. Grishchuk, ``{Amplification of gravitational waves in an istropic
  universe},'' {\em Sov. Phys. JETP} {\bf 40} (1975)
409--415.

\bibitem{Rubakov:1982df}
V.~A. Rubakov, M.~V. Sazhin, and A.~V. Veryaskin, ``{Graviton Creation in the
  Inflationary Universe and the Grand Unification Scale},'' {\em Phys. Lett.}
  {\bf B115} (1982)
189--192.

\bibitem{Gangui:1993tt}
A.~Gangui, F.~Lucchin, S.~Matarrese, and S.~Mollerach, ``{The Three point
  correlation function of the cosmic microwave background in inflationary
  models},'' {\em Astrophys. J.} {\bf 430} (1994) 447--457,
\href{http://arXiv.org/abs/astro-ph/9312033}{astro-ph/9312033}.

\bibitem{Maldacena:2002vr}
J.~M. Maldacena, ``{Non-Gaussian features of primordial fluctuations in single
  field inflationary models},'' {\em JHEP} {\bf 05} (2003) 013,
\href{http://arXiv.org/abs/astro-ph/0210603}{astro-ph/0210603}.

\bibitem{Chen:2006nt}
X.~Chen, M.-x. Huang, S.~Kachru, and G.~Shiu, ``{Observational signatures and
  non-Gaussianities of general single field inflation},'' {\em JCAP} {\bf 0701}
  (2007) 002,
\href{http://arXiv.org/abs/hep-th/0605045}{hep-th/0605045}.

\bibitem{Page:2006hz}
{\bf WMAP} Collaboration, L.~Page {\em et al.}, ``{Three year Wilkinson
  Microwave Anisotropy Probe (WMAP) observations: Polarization analysis},''
  {\em Astrophys. J. Suppl.} {\bf 170} (2007) 335,
\href{http://arXiv.org/abs/astro-ph/0603450}{astro-ph/0603450}.

\bibitem{Komatsu:2008hk}
{\bf WMAP} Collaboration, E.~Komatsu {\em et al.}, ``{Five-Year Wilkinson
  Microwave Anisotropy Probe (WMAP) Observations:Cosmological
  Interpretation},''
\href{http://arXiv.org/abs/0803.0547}{0803.0547}.

\bibitem{Yadav:2007yy}
A.~P.~S. Yadav and B.~D. Wandelt, ``{Detection of primordial non-Gaussianity
  (fNL) in the WMAP 3-year data at above 99.5\% confidence},''
\href{http://arXiv.org/abs/arXiv:0712.1148}{arXiv:0712.1148 [astro-ph]}.

\bibitem{Garriga:1999vw}
J.~Garriga and V.~F. Mukhanov, ``Perturbations in k-inflation,'' {\em Phys.
  Lett.} {\bf B458} (1999) 219--225,
\href{http://arXiv.org/abs/hep-th/9904176}{hep-th/9904176}.

\bibitem{Copeland:1993zn}
E.~J. Copeland, E.~W. Kolb, A.~R. Liddle, and J.~E. Lidsey, ``{Reconstructing
  the inflaton potential: Perturbative reconstruction to second order},'' {\em
  Phys. Rev.} {\bf D49} (1994) 1840--1844,
\href{http://arXiv.org/abs/astro-ph/9308044}{astro-ph/9308044}.

\bibitem{Kolb:1994eu}
E.~W. Kolb, M.~Abney, E.~J. Copeland, A.~R. Liddle, and J.~E. Lidsey,
  ``{Reconstructing the inflaton potential},''
\href{http://arXiv.org/abs/astro-ph/9407021}{astro-ph/9407021}.

\bibitem{Copeland:1993jj}
E.~J. Copeland, E.~W. Kolb, A.~R. Liddle, and J.~E. Lidsey, ``{Reconstructing
  the inflation potential, in principle and in practice},'' {\em Phys. Rev.}
  {\bf D48} (1993) 2529--2547,
\href{http://arXiv.org/abs/hep-ph/9303288}{hep-ph/9303288}.

\bibitem{Lidsey:1995np}
J.~E. Lidsey {\em et al.}, ``{Reconstructing the inflaton potential: An
  overview},'' {\em Rev. Mod. Phys.} {\bf 69} (1997) 373--410,
\href{http://arXiv.org/abs/astro-ph/9508078}{astro-ph/9508078}.

\bibitem{Lesgourgues:2007gp}
J.~Lesgourgues and W.~Valkenburg, ``{New constraints on the observable inflaton
  potential from WMAP and SDSS},'' {\em Phys. Rev.} {\bf D75} (2007) 123519,
\href{http://arXiv.org/abs/astro-ph/0703625}{astro-ph/0703625}.

\bibitem{Lesgourgues:2007aa}
J.~Lesgourgues, A.~A. Starobinsky, and W.~Valkenburg, ``{What do WMAP and SDSS
  really tell about inflation?},'' {\em JCAP} {\bf 0801} (2008) 010,
\href{http://arXiv.org/abs/0710.1630}{0710.1630}.

\bibitem{Hamann:2008pb}
J.~Hamann, J.~Lesgourgues, and W.~Valkenburg, ``{How to constrain inflationary
  parameter space with minimal priors},'' {\em JCAP} {\bf 0804} (2008) 016,
\href{http://arXiv.org/abs/0802.0505}{0802.0505}.

\bibitem{Silverstein:2003hf}
E.~Silverstein and D.~Tong, ``{Scalar speed limits and cosmology: Acceleration
  from D- cceleration},'' {\em Phys. Rev.} {\bf D70} (2004) 103505,
\href{http://arXiv.org/abs/hep-th/0310221}{hep-th/0310221}.

\bibitem{Chen:2004gc}
X.~Chen, ``{Multi-throat brane inflation},'' {\em Phys. Rev.} {\bf D71} (2005)
  063506,
\href{http://arXiv.org/abs/hep-th/0408084}{hep-th/0408084}.

\bibitem{ArmendarizPicon:1999rj}
C.~Armendariz-Picon, T.~Damour, and V.~F. Mukhanov, ``{k-inflation},'' {\em
  Phys. Lett.} {\bf B458} (1999) 209--218,
\href{http://arXiv.org/abs/hep-th/9904075}{hep-th/9904075}.

\bibitem{ArkaniHamed:2003uz}
N.~Arkani-Hamed, P.~Creminelli, S.~Mukohyama, and M.~Zaldarriaga, ``{Ghost
  inflation},'' {\em JCAP} {\bf 0404} (2004) 001,
\href{http://arXiv.org/abs/hep-th/0312100}{hep-th/0312100}.

\bibitem{Peiris:2007gz}
H.~V. Peiris, D.~Baumann, B.~Friedman, and A.~Cooray, ``Phenomenology of
  D-Brane Inflation with General Speed of Sound,''
\href{http://arXiv.org/abs/arXiv:0706.1240}{arXiv:0706.1240 [astro-ph]}.

\bibitem{Bean:2008ga}
R.~Bean, D.~J.~H. Chung, and G.~Geshnizjani, ``{Reconstructing a general
  inflationary action},''
\href{http://arXiv.org/abs/arXiv:0801.0742}{arXiv:0801.0742 [astro-ph]}.

\bibitem{Li:2008qc}
M.~Li, T.~Wang, and Y.~Wang, ``{General Single Field Inflation with Large
  Positive Non- Gaussianity},'' {\em JCAP} {\bf 0803} (2008) 028,
\href{http://arXiv.org/abs/0801.0040}{0801.0040}.

\bibitem{Cline:2006db}
J.~M. Cline and L.~Hoi, ``{Inflationary potential reconstruction for a WMAP
  running power spectrum},'' {\em JCAP} {\bf 0606} (2006) 007,
\href{http://arXiv.org/abs/astro-ph/0603403}{astro-ph/0603403}.

\bibitem{Babich:2004gb}
D.~Babich, P.~Creminelli, and M.~Zaldarriaga, ``{The shape of
  non-Gaussianities},'' {\em JCAP} {\bf 0408} (2004) 009,
\href{http://arXiv.org/abs/astro-ph/0405356}{astro-ph/0405356}.

\bibitem{Lucchin:1984yf}
F.~Lucchin and S.~Matarrese, ``Power Law Inflation,'' {\em Phys. Rev.} {\bf
  D32} (1985)
1316.

\bibitem{Grana:2005jc}
M.~Grana, ``{Flux compactifications in string theory: A comprehensive
  review},'' {\em Phys. Rept.} {\bf 423} (2006) 91--158,
\href{http://arXiv.org/abs/hep-th/0509003}{hep-th/0509003}.

\bibitem{Giddings:2001yu}
S.~B. Giddings, S.~Kachru, and J.~Polchinski, ``{Hierarchies from fluxes in
  string compactifications},'' {\em Phys. Rev.} {\bf D66} (2002) 106006,
\href{http://arXiv.org/abs/hep-th/0105097}{hep-th/0105097}.

\bibitem{Klebanov:2000hb}
I.~R. Klebanov and M.~J. Strassler, ``{Supergravity and a confining gauge
  theory: Duality cascades and chiSB-resolution of naked singularities},'' {\em
  JHEP} {\bf 08} (2000) 052,
\href{http://arXiv.org/abs/hep-th/0007191}{hep-th/0007191}.

\bibitem{Easson:2007dh}
D.~A. Easson, R.~Gregory, D.~F. Mota, G.~Tasinato, and I.~Zavala,
  ``{Spinflation},''
\href{http://arXiv.org/abs/arXiv:0709.2666}{arXiv:0709.2666 [hep-th]}.

\bibitem{Brandenberger:2007ca}
R.~H. Brandenberger, A.~R. Frey, and L.~C. Lorenz, ``{Entropy Fluctuations in
  Brane Inflation Models},''
\href{http://arXiv.org/abs/arXiv:0712.2178}{arXiv:0712.2178 [hep-th]}.

\bibitem{Randall:1999ee}
L.~Randall and R.~Sundrum, ``{A large mass hierarchy from a small extra
  dimension},'' {\em Phys. Rev. Lett.} {\bf 83} (1999) 3370--3373,
\href{http://arXiv.org/abs/hep-ph/9905221}{hep-ph/9905221}.

\bibitem{Alishahiha:2004eh}
M.~Alishahiha, E.~Silverstein, and D.~Tong, ``{DBI in the sky},'' {\em Phys.
  Rev.} {\bf D70} (2004) 123505,
\href{http://arXiv.org/abs/hep-th/0404084}{hep-th/0404084}.

\bibitem{Chen:2005ad}
X.~Chen, ``{Inflation from warped space},'' {\em JHEP} {\bf 08} (2005) 045,
\href{http://arXiv.org/abs/hep-th/0501184}{hep-th/0501184}.

\bibitem{Barnaby:2004gg}
N.~Barnaby, C.~P. Burgess, and J.~M. Cline, ``{Warped reheating in
  brane-antibrane inflation},'' {\em JCAP} {\bf 0504} (2005) 007,
\href{http://arXiv.org/abs/hep-th/0412040}{hep-th/0412040}.

\bibitem{Chimento:2007es}
L.~P. Chimento and R.~Lazkoz, ``{Bridging geometries and potentials in DBI
  cosmologies},''
\href{http://arXiv.org/abs/arXiv:0711.0712}{arXiv:0711.0712 [hep-th]}.

\bibitem{Baumann:2006cd}
D.~Baumann and L.~McAllister, ``A microscopic limit on gravitational waves from
  D-brane inflation,'' {\em Phys. Rev.} {\bf D75} (2007) 123508,
\href{http://arXiv.org/abs/hep-th/0610285}{hep-th/0610285}.

\bibitem{Becker:2007ui}
M.~Becker, L.~Leblond, and S.~E. Shandera, ``{Inflation from Wrapped Branes},''
  {\em Phys. Rev.} {\bf D76} (2007) 123516,
\href{http://arXiv.org/abs/arXiv:0709.1170}{arXiv:0709.1170 [hep-th]}.

\bibitem{Kobayashi:2007hm}
T.~Kobayashi, S.~Mukohyama, and S.~Kinoshita, ``{Constraints on Wrapped DBI
  Inflation in a Warped Throat},''
\href{http://arXiv.org/abs/arXiv:0708.4285}{arXiv:0708.4285 [hep-th]}.

\bibitem{Thomas:2007sj}
S.~Thomas and J.~Ward, ``{IR Inflation from Multiple Branes},'' {\em Phys.
  Rev.} {\bf D76} (2007) 023509,
\href{http://arXiv.org/abs/hep-th/0702229}{hep-th/0702229}.

\bibitem{Bean:2007eh}
R.~Bean, X.~Chen, H.~V. Peiris, and J.~Xu, ``{Comparing Infrared
  Dirac-Born-Infeld Brane Inflation to Observations},'' {\em Phys. Rev.} {\bf
  D77} (2008) 023527,
\href{http://arXiv.org/abs/arXiv:0710.1812}{arXiv:0710.1812 [hep-th]}.

\end{thebibliography}\endgroup

\end{document}